\documentclass[12pt]{article}
\usepackage{amsmath}
\usepackage{graphicx}
\usepackage{natbib}
\usepackage{url} 
\usepackage{xr-hyper}
\usepackage{hyperref}
\usepackage{lscape}
\usepackage{rotating}

\usepackage{multirow}
\usepackage{ltablex}
\usepackage{ragged2e}
\usepackage{booktabs}

\usepackage{bm}
\usepackage[mathscr]{eucal}

\hypersetup{
    colorlinks = true,
    citecolor = {black}, 
    urlcolor = {blue}, 
    linkcolor = {black}, 
    filecolor={black}
}

\usepackage{helvet}
\usepackage{color}
\usepackage{boxedminipage}
\usepackage{amsfonts}
\usepackage{verbatim}
\usepackage{placeins}

\long\def\gobble#1{}

\usepackage{afterpage,wrapfig} 
\def \IR{\hbox{{\rm I}\kern-.2em\hbox{{\rm R}}}}

\newcommand{\bmone}{\mbox{\bf 1}}

\usepackage{tensor}

\newcommand{\image}[2]{\includegraphics[#1]{#2}}

\defcitealias{KDHS2014}{KDHS, 2014}
\defcitealias{KenyaCensus:2009}{Kenya National Bureau Of Statistics, 2014}

\newcommand{\blind}{0}

\usepackage[T1]{fontenc}
\usepackage{newpxmath}

\usepackage{subcaption}
\begin{document}

\def\spacingset#1{\renewcommand{\baselinestretch}%
{#1}\small\normalsize} \spacingset{1}


\if0\blind
{
  \title{\bf Bayesian Multiresolution Modeling of Georeferenced Data}
  \author{John Paige \thanks{
    John Paige was supported by The National Science Foundation Graduate Research Fellowship Program under award DGE-1256082, and Jon Wakefield was supported by the National Institutes of Health under award R01CAO95994.}\hspace{.2cm}\\
    Department of Statistics, University of Washington,\\
    Geir-Arne Fuglstad \\
    Department of Mathematical Sciences, NTNU, \\
    Andrea Riebler \\
    Department of Mathematical Sciences, NTNU, \\
    and Jon Wakefield \textsuperscript{*} \\
    Departments of Statistics and Biostatistics, University of Washington}
  \maketitle
} \fi

\if1\blind
{
  \bigskip
  \bigskip
  \bigskip
  \begin{center}
    {\LARGE\bf Bayesian Multiresolution Modeling of Georeferenced Data}
\end{center}
  \medskip
} \fi

\bigskip
\begin{abstract}
Current implementations of multiresolution methods are limited in terms of possible types of responses and approaches to inference. We provide a multiresolution approach for spatial analysis of non-Gaussian responses using latent Gaussian models and Bayesian inference via integrated nested Laplace approximation (INLA). The approach builds on `LatticeKrig', but uses a reparameterization of the model parameters that is intuitive and interpretable so that modeling and prior selection can be guided by expert knowledge about the different spatial scales at which dependence acts. The priors can be used to make inference robust and integration over model parameters allows for more accurate posterior estimates of uncertainty.

The extended LatticeKrig (ELK) model is compared to a standard implementation of LatticeKrig (LK), and a standard Matérn model, and we find modest improvement in spatial oversmoothing and prediction for the ELK model for counts of secondary education completion for women in Kenya collected in the 2014 Kenya demographic health survey. Through a simulation study with Gaussian responses and a realistic mix of short and long scale dependencies, we demonstrate that the differences between the three approaches for prediction increases with distance to nearest observation.

\end{abstract}

\noindent%
\vfill

\newpage
\spacingset{1.5} 
\section{Introduction}
\label{sec:intro}

The increasing size and complexity of spatial point datasets 
in fields such as climate sciences, public health, ecology, and
social sciences have been concurrent with methodological developments in spatial statistics. 
While there are currently a host of methods available for handling inference with ``big data'' using traditional spatial models
\citep{heaton2019case}, there has been less focus on accessible tools for more complex spatial 
dependence structures. In the context of multi-resolution spatial modeling, recent developments
are the LatticeKrig (LK) model \citep{nychka:etal:15} with the associated \texttt{R} package \texttt{LatticeKrig} \citep{LatticeKrigPackage}, 
and the multi-resolution approximation (M-RA) model \citep{katzfuss2017multi} with its implementation
in the \texttt{R} package \texttt{GPvecchia} \citep{katzfuss2020general,katzfuss2018vecchia,zilber2019vecchia}. However,
to the best of our knowledge there exist no Bayesian implementations of LK or M-RA allowing for non-Gaussian responses; \texttt{LatticeKrig}
is limited to Gaussian responses as well, and \texttt{GPvecchia} allows general exponential families for the responses.

The most common approach to spatial modeling is to use parametric classes of spatial covariance functions with 
interpretable parameters such as the
Mat\'{e}rn family.  Depending on its
smoothness parameter $\nu$, the Mat\'{e}rn covariance class includes both exponential
and Gaussian covariance functions. However, in practice, the 
smoothness parameter is commonly fixed at a small number, in part due to
the difficulty in estimating this parameter, and the computational benefit of having one fewer parameter \citep{stein:99}. 
It is known that, under infill asymptotics, it is the behavior of the Mat\'{e}rn covariance function at short spatial scales 
that most determines the likelihood and pointwise predictions
\citep[Ch. 3]{stein:99}. This means that while short scale behavior of the Mat\'{e}rm covariance 
may be fit accurately, long scale correlations in the data will often not be accurately 
reproduced by the fit model.  However, as we later show in the simulation study, 
long range correlations become increasingly important
when making predictions far from observations. Additionally, we show in Appendix
A that for areal predictions, errors in 
the covariances at spatial scales close to the average radius of the areas 
affect the uncertainty of those areal predictions the
most, suggesting that long scale correlations are especially relevant when calculating 
the uncertainty of areal averages for large areas.

The difficulty in identifying spatial model parameters
makes it especially important to integrate over uncertainty 
when calculating predictive uncertainty. In a frequentist setting the
bootstrap can be applied, but it relies on asymptotics and is
computationally expensive since it requires the model to be refit many times \citep{sjostedt2003bootstrap}.
\citet{handcock:stern:93} and \citet[Ch. 3.7]{gelfand2010handbook}
recommend using Bayesian inference in spatial statistics due to the importance 
of accounting for uncertain covariance structure. However,
Markov Chain Monte Carlo (MCMC) techniques are often difficult to
implement with long running times and large memory requirements, especially with large numbers of
observations \citep{filippone:etal:13}. Detailed output diagnostics are also necessary to assess
convergence. 

As such, the key limitation in providing Bayesian inference for multiresolution
spatial models is the computational complexity involved. In this paper, we propose
to take advantage of the deterministic algorithm for Bayesian inference
based on Integrated Nested Laplace Approximations (INLA) \citep{rue:etal:09}. 
LK uses different layers of compact basis functions together with an
associated sparse precision matrix, and fits directly into the INLA framework
of latent Gaussian models. We provide an implementation using the \texttt{R} package
\texttt{INLA}, which permits fast and accurate
estimation of posterior marginal densities provided that the number of
parameters is not too big (typically 2 to 5, but not exceeding 20
\citep{rue-etal-2017}). This extended version of LK is termed
extended LatticeKrig (ELK). A key change from the original LK formulation is a reparametrization
that improves interpretability and facilitates modeling and prior selection. Furthermore, the
\texttt{INLA} implementation means that the ELK spatial model can be fit jointly with other random effects such as models for temporal trends or nonlinear covariate effects, handle non-Gaussian responses,
integrate over parameter uncertainty, and incorporate prior
knowledge through expert knowledge and/or for the purpose of robustness.

We will contrast ELK to traditional spatial models using the stochastic
partial differential equation (SPDE) approach \citep{Lindgren:etal:11} as implemented in
\texttt{INLA} to
permit fast Bayesian approximate inference for latent Gaussian models
where the traditional Mat\'{e}rn
covariance function is used for spatial modeling \citep{lindgren:rue:15}. In this context, the SPDE
approach is only one choice among many others for making the computations possible:
employing low rank
covariance matrices
\citep{cressie:johannesson:08,banerjee:etal:08,finley:etal:09}, sparse
covariance matrices
\citep{knorr2000bayesian,sang2012full,konomi2014adaptive,neelon2014multivariate,furrer2006covariance,hirano2013covariance},
sparse precision matrices
\citep{nychka:etal:15,katzfuss2017multi,katzfuss2017parallel,Lindgren:etal:11,datta2016hierarchical,datta2016nearest,guinness2019spectral,guinness2017circulant},
or algorithmic approaches
\citep{gerber2018predicting,guhaniyogi2018meta,gramacy2015local}.

In Section \ref{sec:motivatingExample} we introduce the main application
on prevalence of secondary education for women in
Kenya that motivated this work.  In Section \ref{sec:methods} we describe
LK and ELK. We evaluate ELK, LK, and a SPDE model in a
simulation scenario when fit
to random fields with mixtures of short and long-range correlations in
Section \ref{sec:simulationStudy}. In
Section \ref{sec:application} the ELK and SPDE
models are applied to the real data introduced in Section
\ref{sec:motivatingExample}, and their predictive performance is assessed. Section
\ref{sec:conclusion} concludes this work with a discussion.

\section{Motivating application}
\label{sec:motivatingExample}

\begin{figure}
\centering
\subcaptionbox{Urbanicity}[3.17in]{\image{width=3.17in}{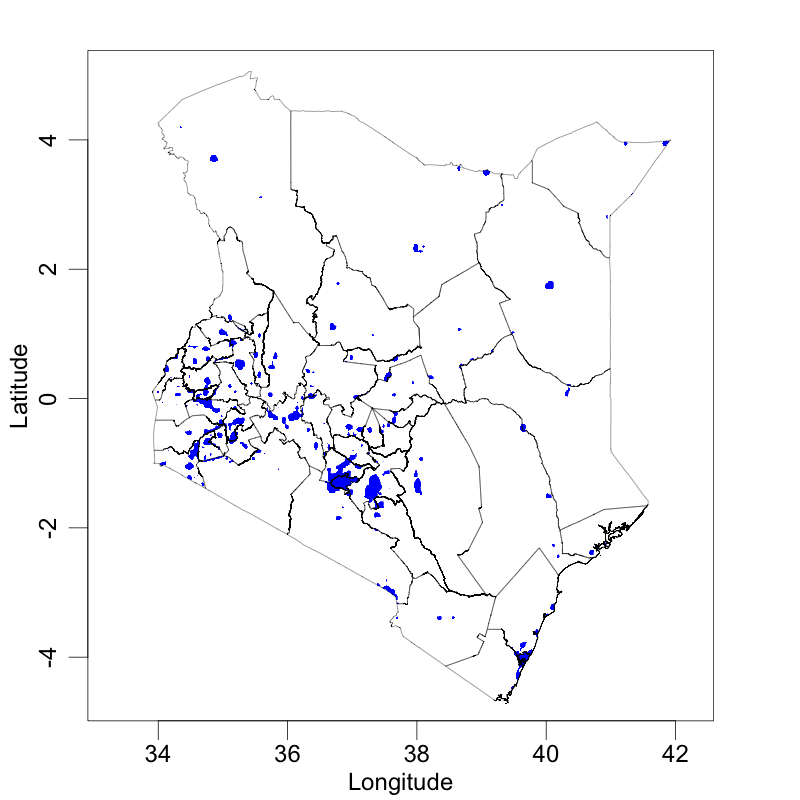}} \subcaptionbox{Secondary education completion}[3.17in]{\image{width=3.17in}{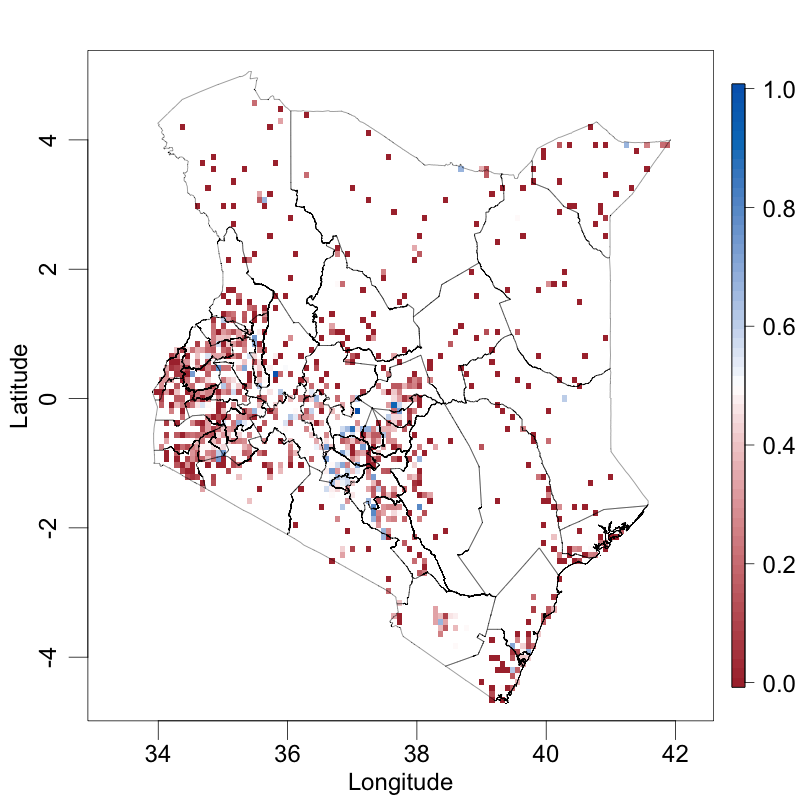}}
\caption{(a) Map of binary urbanicity classification in Kenya, and (b) 2014 empirical proportion of women aged 20-29 in Kenya that completed their secondary education.}
\label{fig:dataset}
\end{figure}


Sustainable Development Goal (SDG) 4 \citep{sdgsWeb} calls for improvements in 
secondary education to the point where everyone can complete their secondary education by 2030 
regardless of their gender or the place where they live. 
Reliable spatial estimates of secondary education completion for
young women are of particular importance to SDG 4. Yet in 
many developing countries, estimates of secondary education completion rely on
complex, multistage household surveys
\citep{li:etal:19,wagner:etal:18} such as demographic health surveys
(DHS) \citep{dhs}, multiple indicator cluster survey (MICS)
\citep{MICS}, AIDS indicator surveys (AIS) \citep{AIS}, and living
standard measurement surveys (LSMS) \citep{LSMS}.

Often, these household surveys are stratified by administration area
and urbanicity; see for instance \citet{samplingManualDHS}.  However, the classifications
of urban or rural for the sampled clusters 
was made at the time of the last census, which
at best takes place every 10 years, and the specific continuous
spatial classifications of urbanicity used in the censuses are
generally not made publicly available. This forces modelers to
either ignore urbanicity or 
assume that the classification remains accurate over large time spans, and to
estimate urbanicity for unobserved locations based on proxy data
such as population density
\citep{paige2020design,wakefield:etal:19}. Because administrative areas are relevant for 
stratification in household surveys, and also since household surveys are often 
used to calculate population averages in administrative areas for policymakers, 
any spatial model used in this context must be able to simultaneously produce accurate 
averages in areas of varying size. Such models will therefore 
need to accurately estimate correlations at all spatial scales relevant for the sizes of the 
areas over which averages are calculated, and account for the uncertainty in those 
correlation estimates.


In this work we consider prevalence of secondary education completion
for young women in Kenya in 2014. Data are obtained from the 2014
Kenya DHS \citepalias{KDHS2014} consisting of 1,612 clusters, each with official urban/rural
designations, and age and educational achievement information for the
sampled women within the cluster. The modeled response is the number
of women aged 20-29 that have completed their secondary
education. \citet{paige2020design} found that there are large 
differences in secondary education completion between urban and rural
areas, and Figure \ref{fig:dataset} shows that urbanicity changes
sharply over short spatial scales. This motivates the development of
spatial models that can include spatial dependence at widely different
spatial scales. We revisit this dataset in Section \ref{sec:application} to
explore the sensitivity of spatial analysis to the inclusion or
non-inclusion of an urban covariate, and the degree to which the ELK model
can guard against spatial oversmoothing when urbanicity is not included
and how well short and long scale correlations are captured when urbanicity
is included.

\section{Methods}
\label{sec:methods}

\subsection{Background on LatticeKrig}
\label{sec:LK}
\citet{nychka:etal:15} introduced LK as a computationally efficient method for spatial modeling the stochastic process
$Y = \{y(\boldsymbol{x}) : \boldsymbol{x} \in \mathcal{D} \}$ for spatial domain $\mathcal{D}$ measured at observation locations $\boldsymbol{x}_1, \boldsymbol{x}_2, \ldots\boldsymbol{x}_n\in\mathbb{R}^2$. The observation model was assumed
to be Gaussian, with $y(\boldsymbol{x}_i)|\eta_i, \sigma_\mathrm{N}^2\sim\mathcal{N}(\eta_i, \sigma_\mathrm{N}^2)$, $i = 1, \ldots, n$, where
$\boldsymbol{\eta} = (\eta_1, \eta_2, \ldots, \eta_n)$ were the linear predictors and 
$\sigma_\mathrm{N}^2$ was the nugget variance. The linear predictors were assumed to follow
 a linear model 
$\boldsymbol{\eta} = \mathbf{Z} \boldsymbol{\beta} + \boldsymbol{u}$, where
$\mathbf{Z}$ is a $n\times p$ matrix where each column specifies a covariate, $\boldsymbol{\beta} = (\beta_1, \beta_2, \ldots, \beta_p)^T$ is a vector
 containing the coefficients associated with the covariates, and
$\boldsymbol{u} = (u(\boldsymbol{x}_1), u(\boldsymbol{x}_2),\ldots, u(\boldsymbol{x}_n))$ are
the values of the spatial Gaussian random field (GRF) $u$ at the observation locations.

LK is characterized by the decomposition of $u$ into a series of lattices of increasing spatial resolutions over which increasingly fine basis functions are spaced,
\[
  u(\boldsymbol{x}) = \sum_{l=1}^L g_l(\boldsymbol{x}) = \sum_{l=1}^L \sum_{j=1}^{m(l)} c^l_j \phi_{l,j}(\boldsymbol{x}), \quad \boldsymbol{x} \in \mathcal{D}\subset\mathbb{R}^2.
\]
Here $L$ is a fixed, predetermined small number of lattice layers, usually between 2 and 4, and $g_1, ..., g_L$ are a series of smooth spatial functions associated with each lattice and composed of $m(1), ..., m(L)$ basis functions respectively. Each $g_l$ is respectively decomposed into a linear combination of basis functions $ \phi_{l,1}, ..., \phi_{l,m(l)}$ with basis weights $c^l_j$, which are random variables.

\citet{nychka:etal:15} choose radial Wendland basis functions \citep{wendland:95}, which have compact support. The basis functions are represented as $\phi_{l,j}(d) = \phi \left (\frac{d}{2.5 \delta_l} \right )$, where
$ \phi(d) = 
(1-d)^6(35d^2 + 18d + 3)/3$ for $0 \leq d \leq 1$, and 0 otherwise.
Here $\delta_l$ is the layer $l$ lattice cell width, and the factor of
2.5 ensures that the radius of each basis function is 2.5 times the
respective layer lattice cell width. This overlap reduces artifacts in
the predictive spatial means and standard errors \citep{nychka:etal:15}. 

The basis coefficients for each layer respectively follow independent SAR models with mean zero multivariate normal distribution, $\mathbf{c}_l \sim \mbox{MVN}(\mathbf{0},  \alpha_l \sigma_{\mathrm{S}}^2 \mathbf{B}^{-1}_l \mathbf{B}^{-T}_l)$, where $\alpha_l$ determines the proportion of spatial variance $\sigma_{\mathrm{S}}^2$ attributed to layer $l$ with $\sum_{l=1}^L  \alpha_l = 1$, and $\mathbf{B}_l$ is an autoregression matrix for layer $l$ with elements $4 + \kappa_l^2$ on the diagonal and up to four additional non-zero elements on each row corresponding to each neighbor, and with values of $-1$.
As described in \citet{Lindgren:etal:11}, each layer $l$ approximates a Gaussian process with Mat\'{e}rn covariance function having smoothness $\nu=1$ and effective spatial range approximately $\rho_l \equiv \sqrt{8} \delta_l / \kappa_l$. Note that \citet{nychka:etal:15} achieves the desired spatial variance
$\sigma_{\mathrm{S}}^2$ in each point by numerical normalization of the covariance matrix. The interpretation
of the $\alpha_l$ as proportion of variance attributed to layer $l$ is not exact as the marginal
variance of the different layers will vary depending on the values of $\kappa_l$.

Let $\mathbf{A}_l$ be the $n \times m(l)$ regression matrix from the basis coefficients for layer $l$ to the basis function values at the coordinates of the observations so that $(\mathbf{A}_l)_{i,j} = c^l_j \phi_{l,j}(\boldsymbol{x}_i)$. We can then write the regression matrix from all basis coefficients to the values of the basis functions at the observation locations as $\mathbf{A} = (\mathbf{A}_1 \ ... \ \mathbf{A}_L)$ so that $\boldsymbol{u} = \mathbf{A}\boldsymbol{c}$, where $\boldsymbol{c} = (\boldsymbol{c}_1^T \ ... \ \boldsymbol{c}_L^T)^T$. This means that the linear predictor can
be written as $\boldsymbol{\eta} = \mathbf{Z} \boldsymbol{\beta} + \mathbf{A}\boldsymbol{c}$. 

In the above formulation, LK requires $p$ parameters for fixed effects, and $2L+1$ parameters for the covariance including the spatial variance $ \sigma_{\mathrm{S}}^2 $, error variance $ \sigma_N^2$, $L-1$ parameters for the layer weights, and $L$ effective range parameters. It is sometimes assumed for simplicity that $ \kappa_1 =  \kappa_2 = \ldots =  \kappa_L$, in which case the effective range of each layer is controlled exclusively by the layer resolution. Under this assumption, LK requires only $L + 2$ covariance parameters.

\subsection{A Bayesian extension to latent Gaussian models}
\label{sec:ELK}
We make two major additions to the formulation in the previous section: we allow for 
the model to be fit jointly with other structured random effects, and we allow for non-Gaussian responses. The model for the linear predictor is extended to
$\boldsymbol{\eta} = \mathbf{Z} \boldsymbol{\beta} + \mathbf{A} \mathbf{c} + \sum_{i = 1} \mathbf{M}_i \boldsymbol{\gamma}_i$, where the matrices $\mathbf{M}_i$ are fixed and define a mapping to the observations
from random effects collected in the vectors $\boldsymbol{\gamma}_i$ such as temporal trends, space-time interactions, and 
other modeled effects. The vector $\boldsymbol{\gamma}=(\boldsymbol{\gamma}_1^T, \ldots, \boldsymbol{\gamma}_m^T)^T$ is assumed to follow a joint Gaussian
distribution. Denote by $\boldsymbol{\theta}_\mathrm{M}$ 
and $\boldsymbol{\theta}_\mathrm{L}$ the vectors containing all model and family likelihood hyperparameters respectively. 
We can then formulate a latent Gaussian model in three stages. In stage 1, we have conditionally independent observations that may be non-Gaussian with likelihood
$\pi(y(\boldsymbol{x}_i) | \eta_i, \boldsymbol{\theta}_\mathrm{L})$, $i = 1, 2, \ldots, n$. 
In stage 2, the latent model is a joint Gaussian
distribution for $(\boldsymbol{\beta}, \boldsymbol{\gamma}, \boldsymbol{c}) | \boldsymbol{\theta}_\mathrm{M}$. Lastly, in stage 3, we assign a prior $\pi(\boldsymbol{\theta})$, where $\boldsymbol{\theta} = (\boldsymbol{\theta}_\mathrm{M}, \boldsymbol{\theta}_\mathrm{L})$.

To better understand the $\sum_{i = 1} \mathbf{M}_i \boldsymbol{\gamma}_i$ term, and to see why it adds so much generality to ELK, we could consider the relatively simple example of modeling a set of $T$ repeated observations of $n$ spatial locations through time points $t=1,\ldots,T$. If our covariates aside from $\beta_0$, the intercept, can be split into one set of covariates changing only in space and one set of covariates changing only in time, we could then model the fixed effects in space and time as $\mathbf{Z}_{\mathrm{S}} \boldsymbol{\beta}_{\mathrm{S}}$ and $\mathbf{Z}_{\mathrm{T}} \boldsymbol{\beta}_{\mathrm{T}}$ respectively for $n\times p_{\mathrm{S}}$ matrix $\mathbf{Z}_{\mathrm{S}}$ and $T \times p_{\mathrm{T}}$ matrix $\mathbf{Z}_{\mathrm{T}}$. Similarly, we might assume that the spatial random effect varied only in space and the temporal random effect varied only in time.  If the temporal trend is AR(1), then we can set $\boldsymbol{\gamma} \sim $ AR(1), for a $T$ dimensional vector $\boldsymbol{\gamma}$. We could then define the model as, $\boldsymbol{ \eta } = \boldsymbol{1}_{nT}\beta_0 + (\boldsymbol{1}_T \otimes \mathbf{Z}_{\mathrm{S}}) \boldsymbol{ \beta }_{\mathrm{S}} + (\mathbf{Z}_{\mathrm{T}} \otimes \boldsymbol{1}_n) \boldsymbol{ \beta }_{\mathrm{T}} + (\boldsymbol{1}_T \otimes \mathbf{A}) \mathbf{c} + (\mathbf{I}_T \otimes \boldsymbol{1}_n) \boldsymbol{  \gamma }$, where `$\otimes$' represents the Kronecker product, and $\mathbf{I}_T$ is a $T \times T$ identity matrix so that $\mathbf{M}=\mathbf{I}_T \otimes \boldsymbol{1}_n$ adds the coefficients of $\boldsymbol{ \gamma }$ identically to the coefficients of $\boldsymbol{ \eta }$ associated with the corresponding time point. This model can be fit in the ELK framework. Although not included in this model, interactions between the spatial and temporal effects could be included as well.

The key computational contribution of \citet{rue2009approximate} is the combination of this formulation 
with the INLA approach to make
Bayesian inference for the multiresolution 
latent Gaussian model computationally feasible. The combination is practically achieved
by the implementation of the new model within the \texttt{INLA}
package. We term the extended version of LatticeKrig, with computationally feasible
inference, as extended LatticeKrig (ELK). Our implementation exploits \verb|GMRFLib|-library \citep{rue2001gmrflib} functions for sparse symmetric positive definite matrices based 
on methods described in \citet{rue2005gaussian} when generating the latent coefficient precision and covariance matrices, 
and also precomputes relevant matrices and normalization factors whenever possible.  Details on computations involved in our ELK implementation are given in Appendix \ref{sec:computation}.

To ensure $\sigma_{\mathrm{S}}^2$ can be approximately interpreted as the spatial variance and 
$(\alpha_1, \ldots, \alpha_L)$ as the proportion of spatial variance attributed to
the layers, we normalize separately the SAR processes associated with each layer so that the 
variance of each $g_l$ in the center of the spatial domain is $\alpha_l \cdot \sigma_{\mathrm{S}}^2$.  
This requires the computation of normalization constants $ \omega_1, ...,  \omega_L$. Letting $\mathbf{A}^*_l$ be the $1 \times m(l)$ 
regression row vector that maps the layer $l$ basis coefficients to the value of the basis functions at the center 
of the spatial domain, each $ \omega_l$ can be calculated as: 
\(
\omega_l = (\mathbf{A}^*_l \mathbf{B}^{-1}_l \mathbf{B}^{-T}_l (\mathbf{A}^*_l)^T)^{-1}. 
\)
This is different from LK, since we only normalize the process to have variance $\sigma_{S}^2$ in the center of the domain rather than at every point.  This has the advantage that it is faster computationally, and we find that if the lattice resolutions and buffers are chosen using the method discussed in the following paragraph, then the resulting process has spatial variance close to $\sigma_S^2$ across the whole spatial domain. In order to avoid matrix inversion and quadratic form computations each time $\mathbf{Q}$
is calculated, we precompute the mappings $f_l: \kappa_l \mapsto \omega_l$ using smoothing splines over a reasonable range of the values of $ \kappa_l$.

In LK, the recommended setting for the layer resolutions are the 
relation $ \delta_l = 2^{-(l-1)} \delta_1$, and when this relation is used in ELK under the assumption that $ \kappa_1 = \ldots =  \kappa_L$, we call this the `fixed' model (ELK-F). We propose to also consider
a `tailored' ELK model (ELK-T) with resolutions chosen for capturing variation at different spatial scales and with $ \kappa_l$ parameters allowed to vary for each layer. Since ELK-T allows the $ \kappa_l$ parameters to vary for each layer, it requires $2L$ hyperparameters, whereas ELK-F requires $L + 1$ hyperparameters, although more would be required if other latent effects were included in the $\mathbf{M}\boldsymbol{ \gamma }$ term or for any likelihood family hyperparameters. 
For both models, a conservative guideline is for lattice resolutions to be 
at most a fifth of the effective range of the corresponding layer to avoid lattice artifacts and for accurate interpretation of the 
layer's effective range parameter.  Since correlation lengths near the 
spatial domain diameter are very difficult to identify, we recommend choosing $\delta_1$ to be finer than a fifth of the 
spatial domain diameter, and typically around a twenty fifth of the domain diameter, although the exact choice will depend 
on the context. Figure 8
in Section S.2 in the supplementary material illustrates how the lattices might be arranged for a specific problem. In the figure and in Section \ref{sec:simulationStudy} we use a buffer of 5 cell widths to avoid edge effects due to the zero boundary condition for the 
basis coefficients of each layer. The buffer size can be adjusted depending on the estimated effective correlation 
range for that layer.

Expert knowledge on spatial scales at which dependence is expected could be
used to choose appropriate resolutions in ELK-T. Furthermore, the Bayesian
formulation allows the inclusion of expert knowledge when setting priors for the interpretable parameters. 
For simplicity, we suggest a Dirichlet distribution of order $L$ for the proportion
of variances assigned to each layer, that
is $\boldsymbol{ \alpha } \sim
\mbox{Dirichlet}(a_1, \ldots, a_L)$
with $a_l = 1.5/L$ for $l=1,\ldots,L$ in order to place equal weight in the prior on 
each layer, and to ensure the prior is slightly concave for the sake of identifiability. 
Since the chosen prior concentration parameter is $1.5$, 
the Dirichlet prior is only slightly more concave than the flat Dirichlet distribution that would result if the 
concentration parameter were $1$. On the spatial
and nugget standard deviation we place penalized complexity (PC) priors satisfying $\mathrm{P}(\sigma_{\mathrm{S}}>1) = 0.01$, although this will depend on the context and prior information. See
\citet{simpson:etal:17} for details on PC priors.

We propose setting independent inverse exponential priors for the effective
range in each layer, where the effective range for layer $l$ is computed as
$\rho_l = \sqrt{8} \delta_l / \kappa_l$. For ELK-T, we recommend beginning by placing  
a prior on one layer's effective range, scaling priors for other layer effective range parameters 
proportionally to the lattice grid cell width $ \delta_l$ for ELK-T.  For ELK-F, a single $ \kappa $ 
parameter is estimated so that $ \kappa = \kappa_1 = ... = \kappa_L$, and only one 
effective range parameter requires 
a prior. When placing priors on ELK-F or ELK-T effective ranges in this way, 
a prior on the effective range for one of the layers 
would therefore determine all effective range priors.  Throughout this work, we set 
the median effective range of the coarsest layer at a fifth of the 
spatial domain diameter, determining any other effective range priors accordingly.  
However, the effective range priors can be customized to 
better suit the context as well the expert knowledge of the modeler.

A fully functional proof-of-concept implementation of ELK is freely available on Github at \url{https://github.com/paigejo/LK-INLA}. Since it is implemented in \texttt{R} and not natively in \texttt{C++}, it does not reach its full potential in terms of speed.

 \section{Assessing performance under multiscale dependence}
 \label{sec:simulationStudy}
 
\subsection {Prediction quality measures}
\label{sec:scoringRules}
The different spatial models are compared using three measures of predictive performance:
the root mean square error (RMSE), the continuous rank probability
score (CRPS) \citep{gneiting:raftery:07}, and the empirical coverage of 80\% prediction intervals. We also compared 
each model's runtime including setup, model fitting, predictions, and predictive and covariance parameter uncertainty. 
For each model considered, we calculate the
measures as an average of its values over each held-out observation. We consider
two hold-out schemes: stratified randomized selection and holding out pre-specified regions.
These are discussed in the section on the application. If observations are counts
with denominator $N_i$ for observation $1 \leq i \leq n$, then we
rescale the counts to be empirical proportions with $y_i = y_i^* /
N_i$ for observed count $y_i^*$. 

Unlike RMSE, CRPS is a strictly proper scoring rule, and as such takes into account the
accuracy of the central predictions as well as the calibration of
the uncertainty. Smaller values are preferable. Prediction intervals at the $80\%$ level are
derived from the $0.1$ and $0.9$ quantiles of the
predictive distribution. They are used to compute the prediction
interval empirical coverage. For empirical proportions, and especially for small denominators, 
a fixed prediction interval will generally not provide the correct
coverage even if the predictive distribution is correct due to the
discreteness of the sample space \citep{geyer2005fuzzy}.  We therefore
follow \citet{geyer2005fuzzy} by calculating \textit{fuzzy} coverage
instead, with details given in Appendix S.1. We
have found that fuzzy coverage is much more precise than
non-randomized coverage, allowing us to be sure that observed over- or
undercoverages are due to the accuracy of the predictive uncertainty
rather than the discreteness of the CIs.

 \subsection{Simulation setting}
\begin{figure}
\centering
\subcaptionbox{One realization of the spatial field}[3.15in]{\image{width=3.15in}{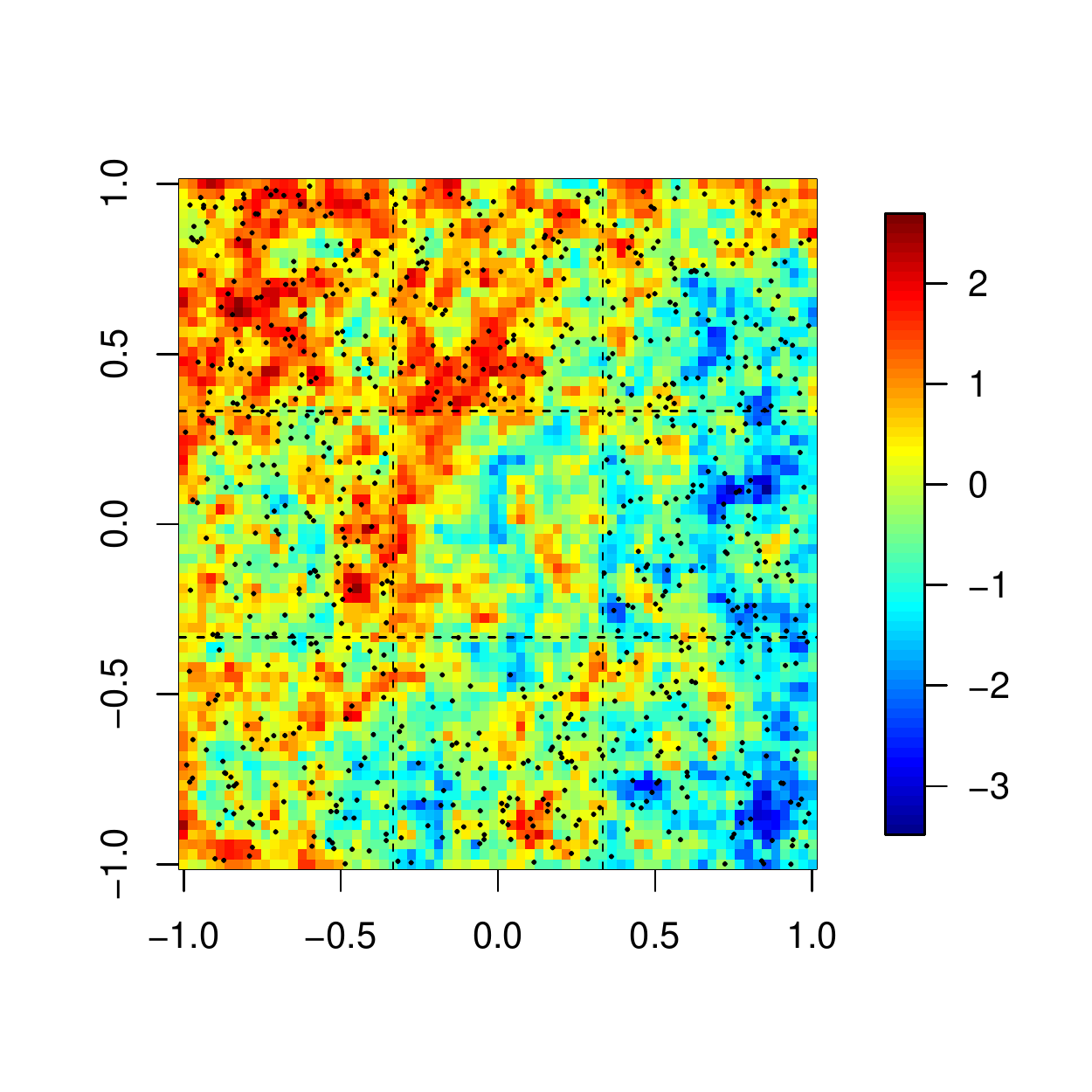}} \subcaptionbox{True and estimated correlation functions}[3.15in]{\image{width=3.15in}{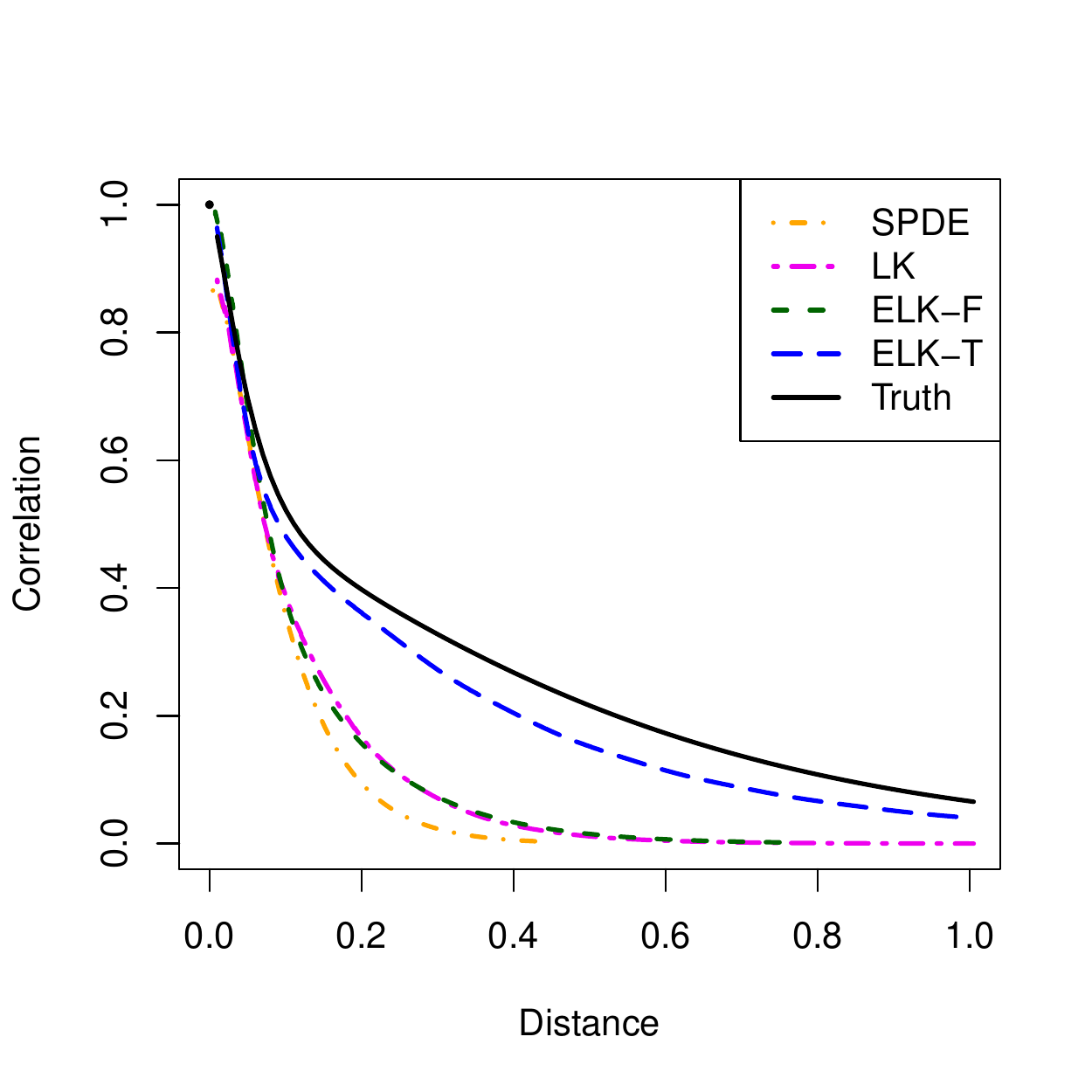}} 
\caption{(a) One of the 100 spatial field realizations. Black dots indicate the 800 observation locations
  and dashed lines indicate the $3 \times 3$ grid used for areal
  predictions (b) True and estimated correlation functions averaged over 100
  realizations.}
\label{fig:exampleSimulation}
\end{figure}

 We first simulate a spatial GRF $u$ on the square $[-1, 1]^2$. 
 The GRF has the covariance function
 $C(d) = 0.5 (C_1^*(d; 0.08) +
C_1^*(d; 0.8))$, where 
$C_1^*(d;\sigma_{\mathrm{S}}^2) = (\sqrt{8}d/ \rho )K_1(\sqrt{8}d/ \rho )$ denotes the Mat\'{e}rn correlation \citep{stein:99} at distance $d$ 
with smoothness $\nu = 1$ and effective spatial range $ \rho $, and where $K_1$ is the modified
Bessel function of the first order and second kind. 
The correlation function is plotted in Figure
\ref{fig:exampleSimulation} along with an example realization.
The domain is then subdivided into 
 a regular $3 \times 3$ grid, and we
draw 800 observations at random locations, $\boldsymbol{x}_1, \boldsymbol{x}_2, \ldots, \boldsymbol{x}_{800}$, in the outer eight grid
cells, but draw no observations within the central grid
cell. We assume the unobserved latent process is $\eta_i = u(\boldsymbol{x}_i)$, and draw 
each observation $Y(\boldsymbol{x}_i)$ from $Y(\boldsymbol{x}_i)|\eta_i\sim \mathcal{N}(\eta_i, 0.1^2)$ for
$i = 1, 2, \ldots, 800$. We fit several models, which we will describe in the next section, to the data,
and generate predictions of the spatial process $Y$ on a fine $70 \times 70$ grid and 
predictions of areal averages of the process $Y$ 
for the nine subdivision areas approximated numerically as averages of the values of $Y$ on the $70 \times 70$ fine grid over each of the 9 areas. The whole procedure is repeated 100 times, and, for each realization, the predictions
are scored in comparison to the truth. We choose to use $Y$ as the process for comparing predictions 
rather than $u$ so that comparison metrics are more similar to cross-validation, where 
only $Y$, and not $u$, is directly observed at the observation locations.

\subsection{Models used in the simulation study}

We use ELK-T with two layers: a grid of $14\times14$ basis knots
and a grid of $126\times126$ knots over the spatial domain (not including the five knot buffer for each layer), 
which results in lattice resolutions of $0.154$ and $0.016$ respectively. In
this case, the coarse and fine scale layers have at least five basis
functions per 0.8 and 0.08 spatial units respectively. Further
we use LK and ELK-F with three layers composed of $14\times14$, $37 \times 37$, and $53\times53$ lattice grids over 
the spatial domain with $0.154$, $0.077$, and $0.038$ resolutions respectively. LK is fit using \texttt{LatticeKrig} 
in \texttt{R}, and for both LK and ELK-F, we use a single layer-independent 
parameter $\kappa$. Additionally, we fit an approximation to the Gaussian process with Mat\'{e}rn 
covariance and smoothness $\nu = 1$ using 
the SPDE approach with \texttt{INLA} \citep{Lindgren:etal:11,lindgren:rue:15}. The mean triangular mesh segment 
length is approximately $0.0064$ within the spatial domain.

This gives in total four models: ELK-T, ELK-F, LK, and SPDE. In all cases
we use a PC prior for the nugget variance satisfying the tail probability $\mathrm{P}(\sigma_{\epsilon} > 1) = 0.01$, and 
for the SPDE model we use the prior derived in \citet{fuglstad2019constructing} on the effective
range and spatial variance. The median effective 
range for the prior is a fifth of the spatial domain diameter, and the spatial
standard deviation again satisfies $\mathrm{P}(\sigma_{\mathrm{S}}>1) = 0.01$. We use PC priors for the 
ELK-T and ELK-F spatial standard deviation also satisfying $\mathrm{P}(\sigma_{\mathrm{S}}>1) = 0.01$, 
and use the effective range priors recommended in Section \ref{sec:ELK}.



\subsection{Results}


For each realization and each of the Bayesian models, we 
generate 1,000 independent samples from the posterior
distribution of $Y$ 
(or conditional distribution in the case of LatticeKrig),
estimate uncertainty in the parameters, and calculate covariance
functions for each of the 100 independent parameter samples. We used 
only 100 parameter samples when generating covariance function draws 
since for each draw the corresponding precision matrix for $u$ must be inverted, 
which is especially computationally intensive for LatticeKrig since it does not 
take advantage of \verb|GMRFLib| library functions for 
factoring sparse symmetric positive definite matrices, and since ELK uses a simplified 
normalization scheme that precomputes normalization factors. 
In the case of LatticeKrig, we use the Hessian of the negative log
likelihood to draw covariance parameter samples.

Figure \ref{fig:exampleSimulation}b)
shows the central correlation function estimate for each of the models together with
the true correlation function. The ELK-T model approximates the true
correlation function over all distances well, while the other
models strongly underestimate the spatial correlation after distance of 0.1, and have negligible 
correlation after distances of approximately 0.5.

\begin{figure}
\centering
\captionsetup[subfigure]{justification=centering}
\subcaptionbox{RMSE vs distance to \\nearest observation}[2.05in]{\image{width=2.05in}{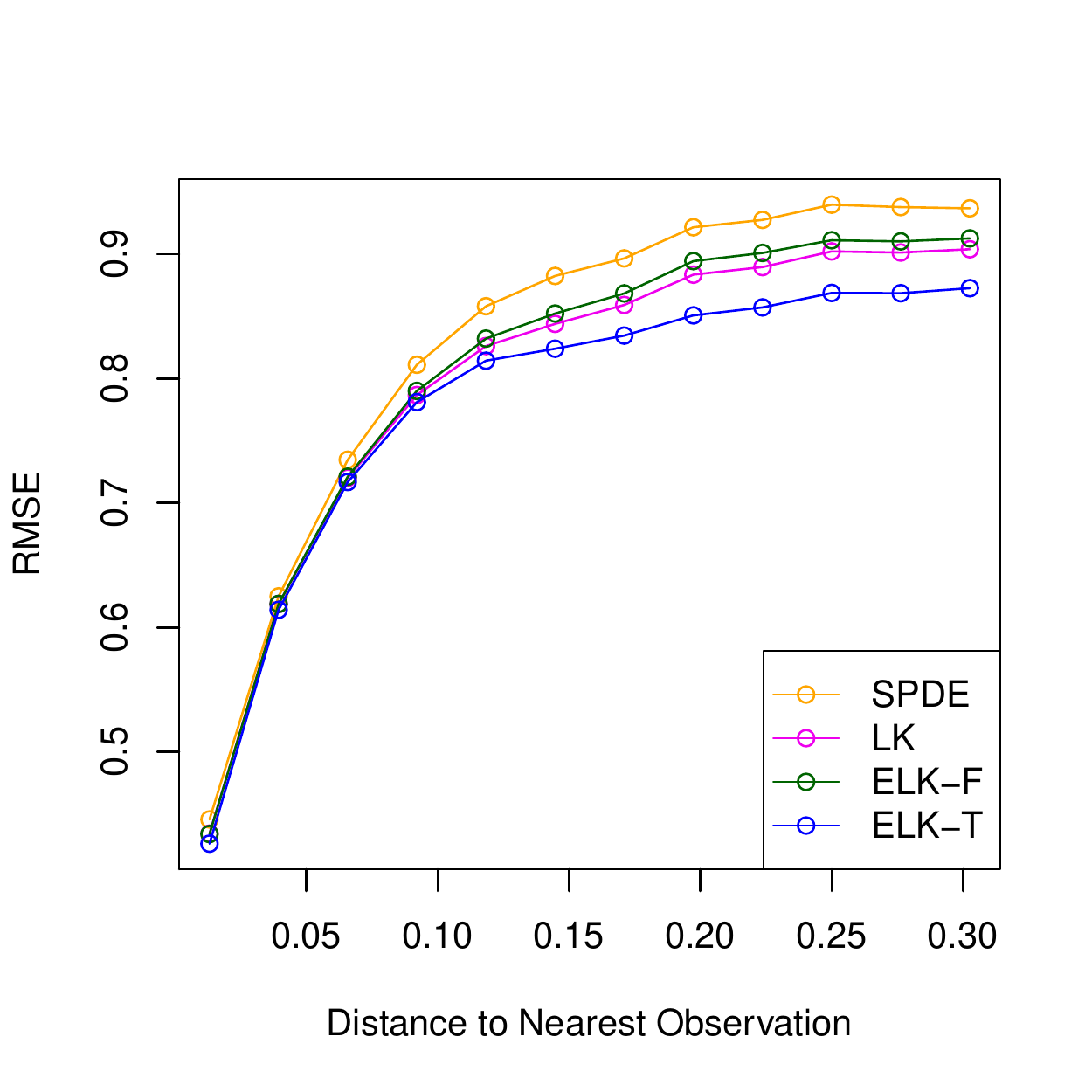}}  \subcaptionbox{CRPS vs distance to \\nearest observation}[2.05in]{\image{width=2.05in}{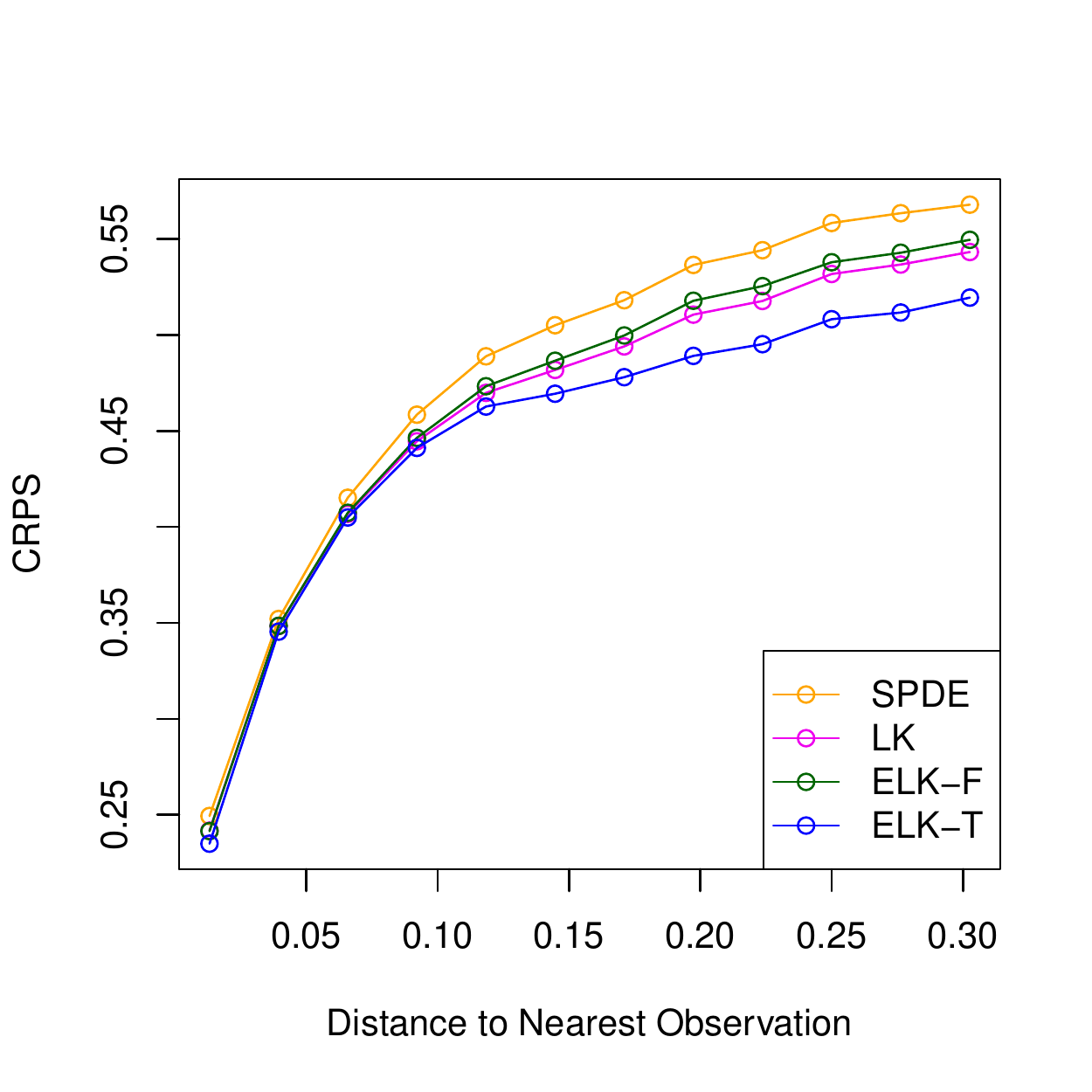}} 
\subcaptionbox{80\% Coverage vs distance to \\nearest observation}[2.05in]{\image{width=2.05in}{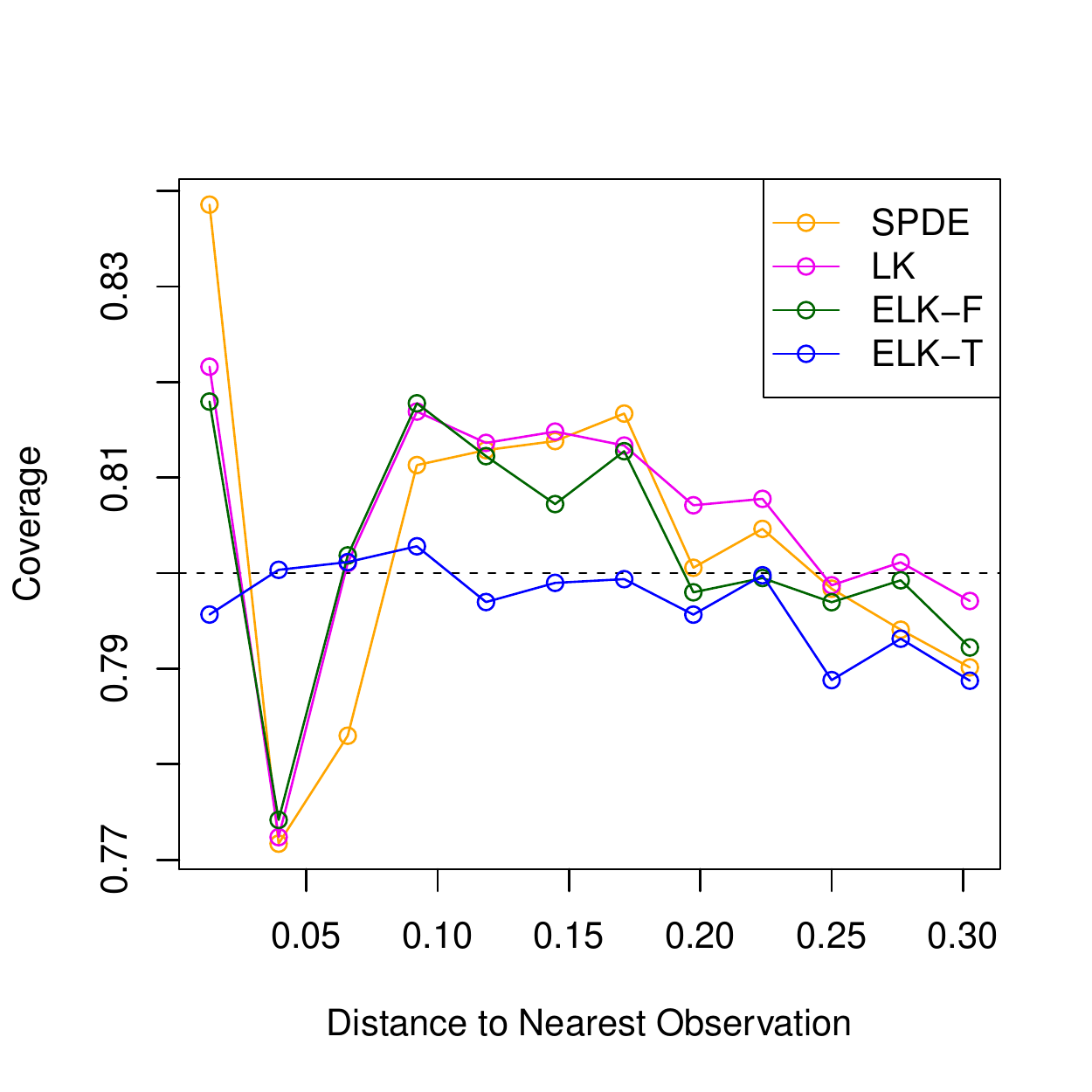}} 
\caption{Scoring rules calculated in bins depending on distance to nearest observation. The scores are averaged over 100 simulations, and include (a) RMSE,  (b) CRPS, and (c) 80\% uncertainty interval coverage.}
\label{fig:illustrationBinnedScores}
\end{figure}

\begin{table}
\centering
\begin{tabular}{lllllll}
\toprule
 & \textbf{RMSE} & \textbf{CRPS} & \textbf{80\% Cvg} & \textbf{Runtime (min.)} \\ 
\bottomrule
\addlinespace[0.3em]
\multicolumn{5}{l}{\textit{\textbf{Pointwise}}}\\

\hspace{1em}SPDE & \textit{0.605} & \textit{0.342} & \textbf{80} &  \textbf{2.0} \\ 
\hspace{1em}LK & 0.594 & 0.334 & \textbf{80} & \textit{51.1} \\ 
\hspace{1em}ELK-F & 0.594 & 0.335 & \textbf{80} &  9.4 \\ 
\hspace{1em}ELK-T & \textbf{0.587} & \textbf{0.329} & \textbf{80} & 12.1 \\ 

\addlinespace[0.3em]
\multicolumn{5}{l}{\textit{\textbf{Areal}}}\\
\hspace{1em}SPDE & \textit{0.137} & \textit{0.056} & \textit{75} &  \textbf{2.0} \\ 
\hspace{1em}LK & 0.121 & 0.051 & 77 & \textit{51.1} \\ 
\hspace{1em}ELK-F & 0.125 & 0.052 & 77 &  9.4 \\ 
\hspace{1em}ELK-T & \textbf{0.108} & \textbf{0.048} & \textbf{79} & 12.1 \\ 
\bottomrule
\end{tabular}
\caption{Scoring rules averaged over 100 simulated realizations and over a regular $70 \times 70$ grid of prediction locations across the entire spatial domain and areally integrated over all nine cells in the $3 \times 3$ regular grid partitioning the domain. Averages are calculated for each of the considered models. \textit{Italics} indicate worse performance, \textbf{boldface} indicates better performance.}
\label{tab:mixtureScores}
\end{table}

Pointwise predictive scoring rules calculated by distance from prediction location
to nearest observation are shown in Figure
\ref{fig:illustrationBinnedScores}. The RMSE
and CRPS of ELK-T are the best in all of the
distance bins. Differences in RMSE and CRPS among the models tend to increase
as the distance to the nearest observation increases, but interestingly the
differentiation is larger in the first bin than in the second bin. 
We believe this is due to the fact that ELK-T
 is able to capture the short range spatial
correlation better than the three other models. 
The differences in RMSE and CRPS values for each
model become increasingly large with longer distance to closest observation,
indicating increasingly differing ability to accurately predict with longer
distances. 

Table \ref{tab:mixtureScores} shows the summarized point and areal prediction
scores. In terms of both pointwise and areal scores, the SPDE predictions have the worst RMSE,
CRPS, and coverage in all cases, although the coverage of all the models marches the nominal 
level of 80\$ in the pointwise case. The coverage of the SPDE model 
is especially poor near the observations, indicating its inability to simultaneously capture short and 
long scale spatial correlations. 
Figure \ref{fig:illustrationBinnedScores} clearly demonstrates that even though
SPDE achieves the correct nominal coverage overall, this is in spite of considerable over- and 
undercoverage depending on how far prediction locations are from the observations. There is also far more variability
in coverage between bins for the SPDE model than ELK-T. 

The runtime for the SPDE model is clearly the best. This is in part due 
to having an implementation that is pre-existing and optimized in the \texttt{INLA} package, whereas
ELK-T and ELK-F were implemented 
manually using the comparatively slow \verb|rgeneric| framework intended for prototyping new models and special cases in \texttt{INLA}. 
However, the fact that the SPDE model requires only two hyperparameters excluding any family likelihood hyperparameters, compared to the four required 
in this case for ELK-F and ELK-T, further improves its computational performance. LK had the longest runtimes in large part 
due to the implementation of the predictive distribution sampling when calculating SEs. Drawing the 1,000 samples took over 
33 minutes on average for LK, whereas drawing the same number of samples for the ELK-F model took under 2 minutes on average, 
and also included sampling over uncertainty in the hyperparameters.

The areal scores in Table \ref{tab:mixtureScores} 
indicate a strong improvement from the
SPDE model to ELK-F, and from ELK-F to ELK-T
in terms of RMSE and CRPS. 
From the SPDE model to ELK-T, pointwise RMSE and CRPS scores improved respectively 
from $0.605$ to $0.587$ ($3.0\%$) and from $0.342$ to $0.329$ ($3.8\%$). However, in the integral prediction
case, RMSE and CRPS scores improved respectively from $0.137$ to $0.108$ ($21\%$) and from $0.056$ to $0.048$ ($14\%$).

Table 5 in Section
S.2 in the supplemental material shows that the improvements in areal predictions
are even larger when considering only the central grid cell, but Table
6 in Section
S.2 shows that there are improvements even when only 
the eight outer grid cells are considered.
In summary, the results of this application show that multi-scale covariance models 
are essential
both for accurate estimation of the covariance structure and for
making predictions when the true covariance function is a mixture of
short range and long range behavior.

\section{Prevalence of secondary education completion}
\label{sec:application}

\subsection{Analysis}
\label{sec:analysis}
We return to the data introduced in Section \ref{sec:motivatingExample}: 
counts of secondary education completion for young women aged 20-29 in Kenya in 2014 using the 2014 Kenya DHS. 
The 2014 Kenya DHS household survey contains responses from individuals sampled from 1,612 clusters in 47 counties, 
each of which except Nairobi and Mombasa (which are both entirely urban) contain both urban and rural strata, making 92 
strata in total. These 47 counties subdivide the 8 geographical provinces in Kenya. 
The response at cluster $c$, conditional on the probability of secondary education completion, $p(\boldsymbol{x}_c)$ at cluster spatial location $\boldsymbol{x}_c$, $c=1, \ldots, 1612$, is modeled as, 
$Y(\boldsymbol{x}_c) | p(\boldsymbol{x}_c) \sim \mbox{Bin}(n_c, p(\boldsymbol{x}_c)),$ where $n_{c}$ is the total number of women aged 20-29 sampled in the cluster. The probability $p(\boldsymbol{x})$ is modeled on logit scale as, 
\begin{equation}
	\eta_c = \log \left( \frac{p(\boldsymbol{x}_c)}{1-p(\boldsymbol{x}_c)} \right) =  \beta_0 + u(\boldsymbol{x}_c) + \beta^{\mbox{\tiny{URB}}} \bmone \{\boldsymbol{x}_c \in U\} + \epsilon_c, \quad c = 1, 2, \ldots, 1612,
	\label{eq:modelB:main}
\end{equation}
with intercept $\beta_0$, spatial random effect $u(\boldsymbol{x}_c)$ with
spatial variance $ \sigma_{\mathrm{S}}^2 $, fixed effect for urban areas
$\beta^{\mbox{\tiny{URB}}}$, and mean zero iid Gaussian cluster random effect
$\epsilon_c$ with variance $ \sigma_\epsilon^2$. The indicator $\bmone
\{\boldsymbol{x}_c \in U\}$ is 1 if $\boldsymbol{x}_c$ is in $U$, the set of
urban areas in Kenya, and 0 otherwise. LK is not applicable due to the binomial likelihood.
We consider four alternatives for $u$:
SPDE\textsubscript{u}/SPDE\textsubscript{U} and
ELK-T\textsubscript{u}/ELK-T\textsubscript{U} models, where `U' and `u'
respectively denote that urban effects are or are not included.

For ELK-T\textsubscript{u} and ELK-T\textsubscript{U}, the coarse lattice layer 
has $37$km
resolution, while the fine layer resolution was set to be $5$km
resolution in order to be able to capture sharp changes from urban
localities to their rural surroundings.
The SPDE model has an average triangular mesh segment length of
approximately 15km across the spatial domain. 
The spatial domain diameter is approximately
$1,445$km, so the prior median effective range was set to be one fifth of that, or $289$km, 
for the SPDE model and for the coarsest layer of the ELK
models. We again place PC priors on the spatial and cluster variance parameters such that
$\mathrm{P}(\sigma_{ \epsilon } > 1) = 0.01$ and $\mathrm{P}(\sigma_{\mathrm{S}} > 1) = 0.01$, 
except now the parameters should be interpreted on logit scale. All covariates except for the intercept are given 
noninformative Gaussian priors with zero mean and $0.001$ precision, and the intercept 
is given an improper $\mbox{Unif}(-\infty, \infty)$ prior.

\begin{figure}
\centering
\subcaptionbox{Covariogram Estimates}[3.17in]{\image{width=3.17in}{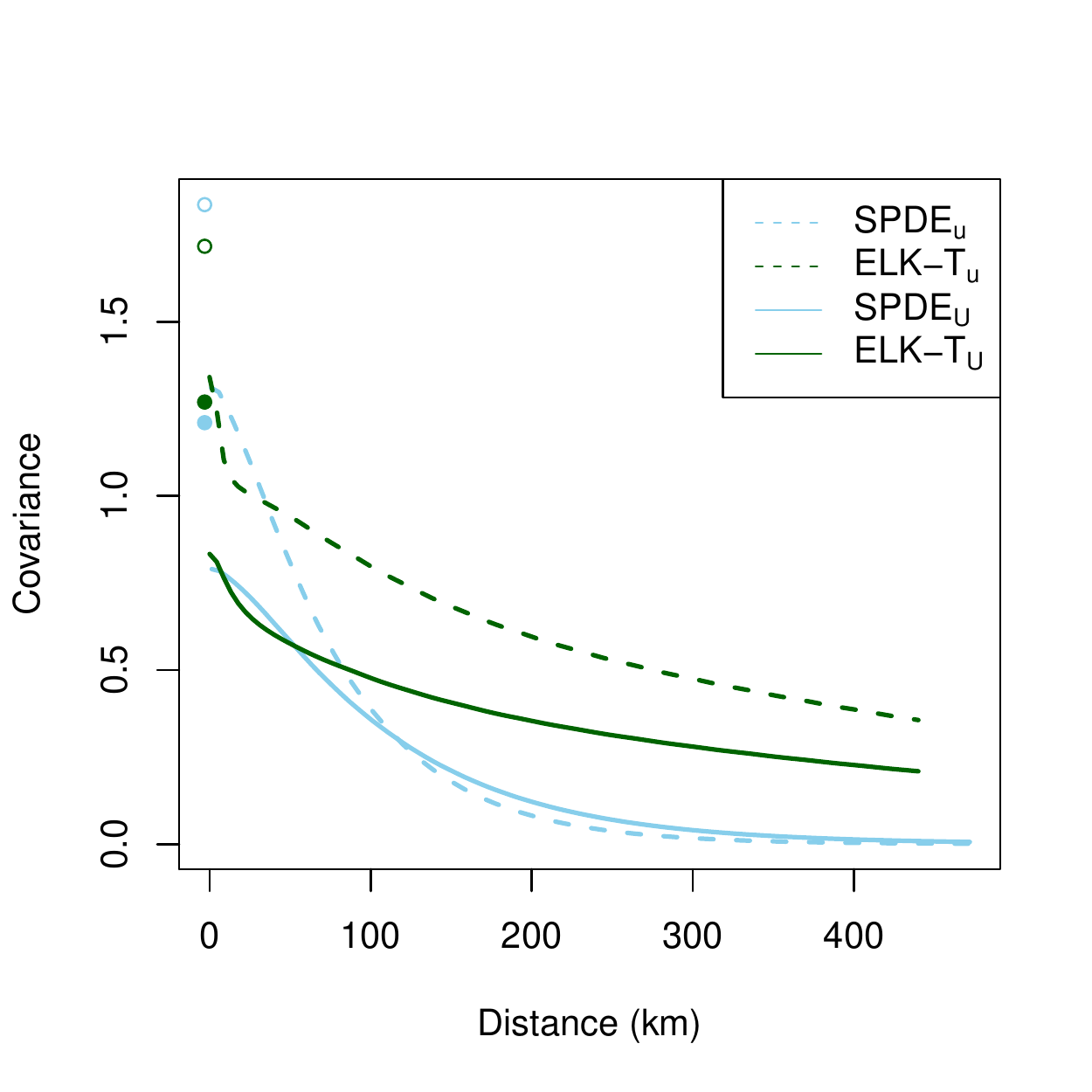}}  
\subcaptionbox{Correlogram Estimates}[3.17in]{\image{width=3.17in}{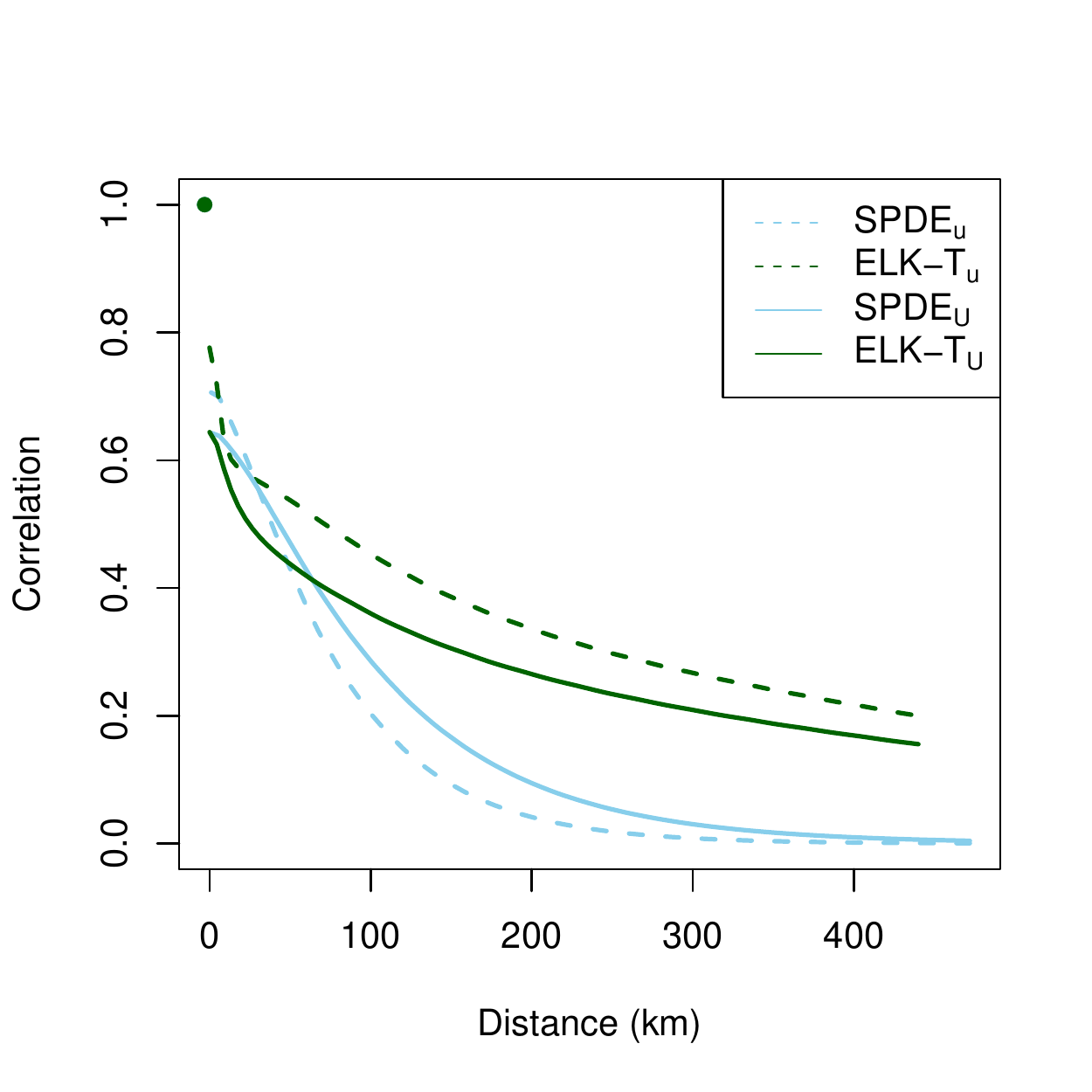}}  
\caption{(a) Spatial covariance, and (b) correlation estimates. The spatial nugget is plotted as the dots at zero distance with the color corresponding to the model given in the legends. Filled dots are plotted for models including urban effects, and unfilled dots are plotted for models without urban effects.}
\label{fig:covarianceCorrelation}
\end{figure}

Central estimates for the correlation and covariance functions of the
fitted models are shown in Figure
\ref{fig:covarianceCorrelation}. Compared to the SPDE models, the
ELK-T models incorporate  more long scale spatial
correlation while also modeling short scale correlations with more
subtlety as shown by their long tailed covariance and correlation
functions with sharp downward trends at small spatial
distances. Including urbanicity as a covariate substantially reduces
the spatial variance for all models, and also reduces the variance of
the spatial nugget. We find that including an urban effect explains
spatial variation at both short and long scales, because sharp changes
due to urban/rural boundaries are accounted for, as well as long scale
correlations across rural regions. We see this effect in the estimated
correlation function of the ELK-T models, where the magnitude of the
relatively sharp downward trend in correlation at small distances
decreases when the urban effect is included, and where the long tail
shortens slightly as well. Since the likelihood under Mat\'{e}rn correlation 
(or Mat\'{e}rn approximations like the SPDE model) is primarily 
affected by short correlation scales, the sharp changes in education due to 
changes in urbanicity rather than the long scale correlations induced by large 
areas being rural drive the correlation function estimate. Hence, including 
the urban effect in the SPDE model removes some of the otherwise unmodeled 
spatial correlation at short spatial scales, increasing  
the estimated effective range. It is worth noting, however, that even with an 
urban effect, the ELK-T\textsubscript{U} model covariance estimates are
still different to those in the SPDE\textsubscript{U} model at both short
and long scales.

\begin{sidewaysfigure}
\centering
\includegraphics[width=\textwidth]{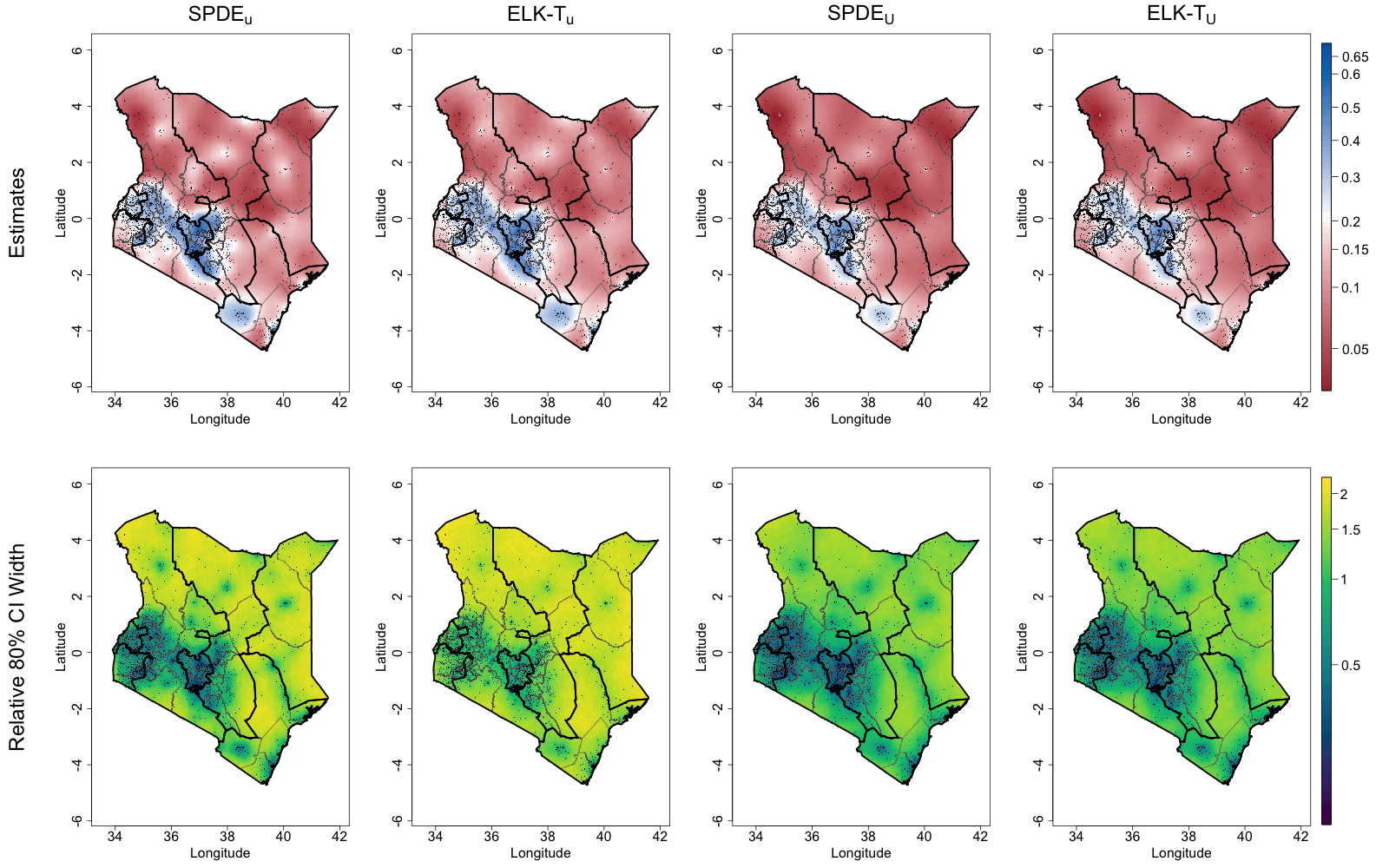}
\caption{Central 5km$ \times $5km pixel level predictions (top row) and relative 80\% credible interval widths (bottom row) of secondary education prevalence for young women in Kenya in 2014. Models with subscript `U' and `u' respectively do and do not include urban effects. Observation locations are plotted as black dots, provinces as thick black lines, and counties as thin gray lines.}
\label{fig:pixelPredictions}
\end{sidewaysfigure}

In Figure \ref{fig:pixelPredictions} we give pixel level predictions at the 5km$ \times $5km resolution of
secondary education prevalence as well as relative credible widths,
which we define as credible widths divided by the corresponding
central estimates. Areal
predictions are created based on aggregation of pixel estimates
weighted by population density as described in Equations (5-6) of \citet{paige2020design}, except 
leaving out cluster effects by setting them to 0 rather than integrating over them as done in Equation (7) of \citet{paige2020design}.
Predictions and relative credible widths aggregated to the county and province levels are shown in
Section S.3 in Figures
9 and 10. Tables of
the county level and province predictions for the models with urban
effects as well summary statistics for the model parameters are given in
S.3 in Tables
7-9.

\begin{figure}
\centering
\image{width=1\textwidth}{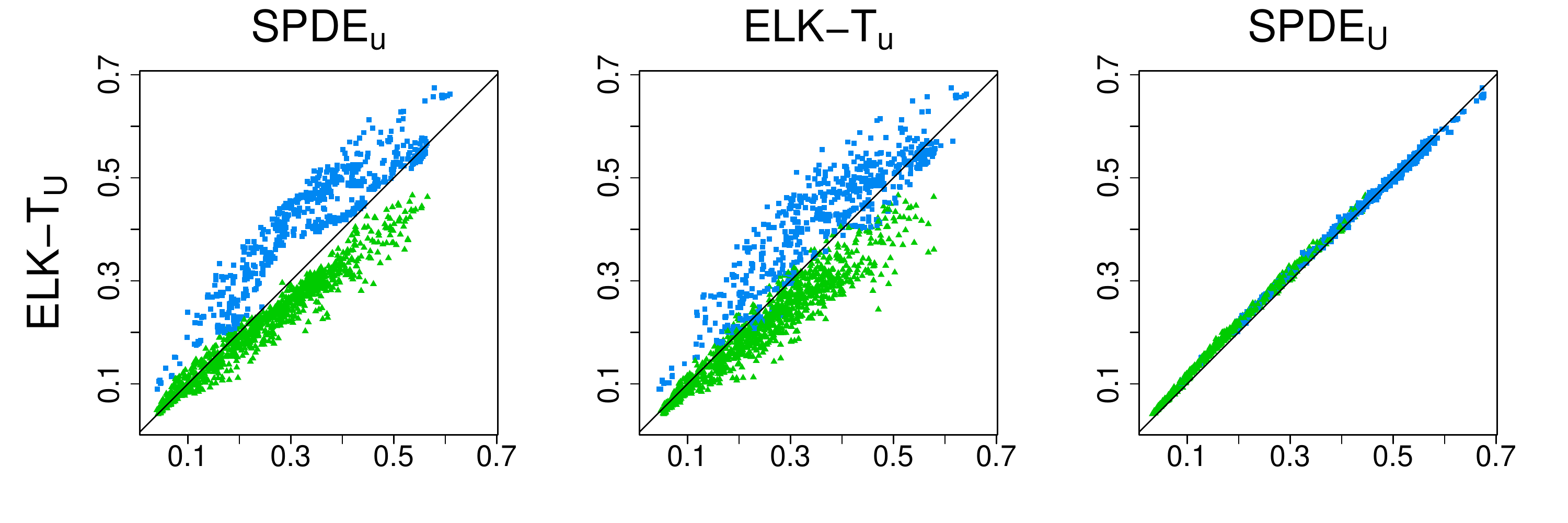}
\caption{Pair plot of the cluster level estimates comparing the considered models' estimates of secondary education prevalence to the ELK-T\textsubscript{U}. The `\protect\includegraphics[height=12pt]{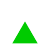}' symbols are rural clusters, while `\protect\includegraphics[height=12pt]{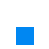}' symbols are urban clusters.}
\label{fig:pairPlot}
\end{figure}

The pixel level predictions show nearly indistinguishable differences
in predictions and uncertainties between the SPDE\textsubscript{U} and
ELK-T\textsubscript{U} models, but much more significant differences in
the predictions between the SPDE\textsubscript{u} and
ELK-T\textsubscript{u} models. In particular, the ELK-T\textsubscript{u}
model shows reduced spatial oversmoothing near urban areas, and higher
uncertainties overall. These uncertainties reflect
that an important confounder in urbanicity is not included as a
covariate. The reduction in oversmoothing is especially noticeable in
the north and east counties with large rural areas and spatially
concentrated urban areas, although there are reductions in
oversmoothing in other areas as well. The differences between the
models without urban effects, and the similarities between the models
with urban effects are further highlighted in the pair plots in Figure
\ref{fig:pairPlot}, which shows the predictions of the
SPDE\textsubscript{u}, ELK-T\textsubscript{u}, and SPDE\textsubscript{U}
models sequentially move towards the predictions of the
ELK-T\textsubscript{U}.

That the SPDE\textsubscript{U} and ELK-T\textsubscript{U} predictions
are essentially indistinguishable lends credence to our predictions by
showing they are robust to modeling assumptions. It also suggests that
there is little identifiable spatial covariance at very short scales
that is not already accounted for by urbanicity, and that the overall
effect of remaining spatial confounders probably
varies smoothly over medium to long spatial scales.

\subsection{Validation}
\label{sec:validation}

We use two different schemes to validate our models: leave one province
out, and stratified, eight-fold cross validation (CV). In the leave
one province out scheme, we calculate scoring rules based on the
predicted distributions of the left out clusters in each of the 8 provinces
consecutively, averaging the scores within each province, and then
averaging the province scores to get the final reported scores. In the
stratified, eight-fold CV, we randomly partition the set of clusters
in each of the 92 strata (47 counties with each except of Nairobi and
Mombasa begin urban and rural) into eight roughly equal sized folds. We make
sure that for a given stratum, the difference between the number of
clusters in each fold is different by at most one, and that which
folds get more clusters than others is random. We choose eight folds
since the smallest stratum has only eight clusters. The two different
validation schemes give an idea of both short and long scale
predictive errors due to the distribution of how far away left out
clusters are from in sample observations. The leave one province out
scheme better identifies long scale errors, and the stratified CV
better identifies short and medium scale errors. The boundaries of the 
8 provinces are plotted in Figure \ref{fig:pixelPredictions} along with 
county boundaries.

\begin{table}
\centering
\begin{tabular}{llrrr}
\toprule
  & \textbf{RMSE} & \textbf{CRPS} & \textbf{80\% Cvg} & \textbf{Width} \\
\bottomrule
\addlinespace[0.3em]
\multicolumn{5}{l}{\textit{\textbf{Leave One Province Out}}}\\
\hspace{2em}SPDE\textsubscript{u} & \em{0.238} & \em{0.129} & 76 & 0.52 \\
\hspace{2em}SPDE\textsubscript{U} & 0.224 & 0.119 & \em{74} & \textbf{0.47} \\
\hspace{2em}ELK-T\textsubscript{u} & 0.234 & 0.125 & \textbf{77} & \em{0.53} \\
\hspace{2em}ELK-T\textsubscript{U} & \textbf{0.223} & \textbf{0.117} & \textbf{77} & 0.49 \\
\addlinespace[0.3em]
\multicolumn{5}{l}{\textit{\textbf{Stratified 8-Fold}}}\\
\hspace{2em}SPDE\textsubscript{u} & \em{0.226} & \em{0.119} & 73 & 0.46 \\
\hspace{2em}SPDE\textsubscript{U} & \textbf{0.218} & 0.114 & \em{72} & \textbf{0.42} \\
\hspace{2em}ELK-T\textsubscript{u} & 0.223 & 0.117 & \textbf{77} & \em{0.49} \\
\hspace{2em}ELK-T\textsubscript{U} & \textbf{0.218} & \textbf{0.113} & 75 & 0.45 \\
\bottomrule
\end{tabular}
\caption{Scoring rules calculated for each model using leave one province out and stratified 8-fold cross validation. Scores are averaged for each province, over urban areas, and over rural areas. \textit{Italics} indicate worse performance, \textbf{boldface} indicates better performance.}
\label{tab:validation}
\end{table}

The results from the leave one province out and the stratified CV are
given in Table \ref{tab:validation}. The ability
of the ELK-T model to account for more flexible spatial covariance
structures than the SPDE model leads to as good or better predictions
as shown by  RMSE, CRPS, and coverage standpoints, although the improvement is
clearly greater when the urban effect is absent in the
model. Improvements were especially obvious in the leave one
province out CV, where long range correlations mattered more, and
relative improvements were greater for CRPS than for RMSE. For leave
one province out CV, RMSE improved by 1.7\% when urban effects were not
included in the SPDE and ELK-T models respectively, and by 0.4\%, while
CRPS improved by 3.1\% when urban effects were not included, and by
1.7\% otherwise. The SPDE\textsubscript{U} model had the worst
coverage with 74\%, and both ELK-T models tied for the best coverage
with 77\%.

\section{Discussion}
\label{sec:conclusion}
The LK approach introduced by \citet{nychka:etal:15}
attempts to address the question of how to
flexibly model spatial covariance at different spatial scales
in a computationally feasible way. However, in a spatial context 
where identifiability is already difficult, spatial confounders and 
the flexibility of LK when layer correlation ranges are allowed 
to independently vary further reduces identifiability. 
In this case, it may be necessary to account for prior information such as expert 
knowledge or to penalize model complexity, and it will certainly be important to integrate over parameter 
and hyperparameter uncertainty. By allowing for this without significant reductions 
in computational performance reductions, ELK's Bayesian framework is a 
valuable extension over standard LK. It is not only more robust, but better 
accounts for multiple levels of uncertainty. Because of this, modelers might be 
less wary of fitting models with more complex covariance structure.

In ELK-T, due to the flexibility in 
choosing layer resolutions and the fact that its 
effective range parameters are fit independently, ELK's Bayesian framework is 
particularly important. We found ELK-T performed much better than 
ELK-F for the simulations we considered and the application since it was better able to 
efficiently model variation at contextually relevant spatial scales. In light 
of this, ELK's use of Bayesian inference is all the more important.

Another advantage of ELK is that it eliminates the assumption of 
Gaussian responses by extending the LK framework to latent
Gaussian models. This allows modeling responses with a diverse set of distributions, such as 
distributions in the exponential family, and even some others such as the 
betabinomial distribution, as long as priors on the latent model components are
Gaussian.  The implementation of ELK in \texttt{INLA}, 
avoids the computational
expense of MCMC when integrating over parameter
uncertainty. Moreover, we show that, computationally, ELK performs
approximately better than LK when uncertainty in the
predictions and covariance parameters is desired.
ELK also has access to the suite of models that can be fit in
\texttt{INLA} such as nonlinear random effects
models for time series or covariates.

It is important to note that ELK-F requires $L+1$ covariance parameters for $L$ layers excluding 
variance parameters of the likelihood family, and ELK-T requires $2L$ 
covariance parameters. Due to the exponential growth in the computation 
time requirements of optimization and integration over hyperparameter uncertainty 
as the number of hyperparameters grow, there is a limit to the number 
of layers for which computation is feasible. It is recommended 
for the number of hyperparameters in \verb|INLA| models to be between 2 
and 5, but certainly not exceeding 20 \citep{rue-etal-2017}. Hence, 
computationally this method should typically use at most 4 layers for ELK-F 
and 2 or 3 layers for ELK-T for likelihoods without extra hyperparameters. 
It is certainly limited to 19 layers for ELK-F 
and 9 layers for ELK-T, which are far more than is necessary for both models. 
In general, for most practical purposes we see little reason to include more than 
3 layers for ELK-T and 5 layers for ELK-F even if computation is feasible due 
to difficulty in model identification and lack of difference in predictive performance, 
although there may be some exceptions to this rule for ELK-F in particular 
since it can only model effective correlation ranges $2^{L-1}$ times larger 
than the range modeled by the finest layer.

In the simulation study, we show that the ability of LK and ELK
to model spatial covariance flexibly can substantially improve
predictive performance at both short and long scales. We find that,
while short scale dependence is most important for point level
predictions near observations, long scale dependence can matter more
when making predictions in data sparse regions, and when making areal
predictions.

When we apply the ELK model to a 2014 Kenya DHS dataset with information on
the prevalence of secondary education for women aged 20-29 in 2014, we
find substantial reductions in spatial oversmoothing relative to a
SPDE model, especially when urbanicity was included as a covariate in
the models. Evidence of short scale spatial confounding was present in
the estimate of the spatial correlation function in the ELK model with
no urban effect, indicating that ELK can make predictions more robust
to spatial confounding as well as be indicative of the spatial scales
at which spatial confounding is occurring.  This in turn can suggest
what variables should be included as covariates, and as an informal check for
spatial confounding. In general, it is very difficult to tell whether an 
unmeasured covariate is confounding results, but ELK provides at 
least a modicum of insurance against this. Since DHS household surveys tend to consist of
clusters that are spatially concentrated in urban areas and sparsely
distributed in rural areas, this is an application that ELK is well
suited for.

Depending on the context, one may choose to select lattice resolutions
that are independent of each other rather than changing by a factor of
two from one layer to the next as in standard LK. In both the
illustrative example and the application, we found that forcing each
consecutive layer to have double the resolution along each dimension
made modeling the fine and long scale changes simultaneously difficult
from a computational perspective due to the number of hyperparameters 
and basis functions required. In such situations, we advocate for
tailoring the resolutions of each lattice to enable them to model a
set of effective ranges of interest.

 \FloatBarrier

\bigskip
\begin{center}
{\large\bf SUPPLEMENTARY MATERIAL}
\end{center}

\begin{description}

\item[Supplements:] Section S.1 in the supplemental material provides details for how we calculate our fuzzy coverage intervals when computing coverage for discrete observations. Section S.2 and S.3 provide additional results for the simulation study and application respectively.

\item[ELK code repository:] Repository with \verb|R| code for fitting the ELK model. Available on Github at: \url{https://github.com/paigejo/LK-INLA}.

\end{description}

\bibliographystyle{apalike}

\bibliography{myBib}

\section*{Appendix A: Relevant Correlation Scales for Spatial Integration}
\label{sec:correlationScales}
\begin{figure}
\centering
\image{width=3.5in}{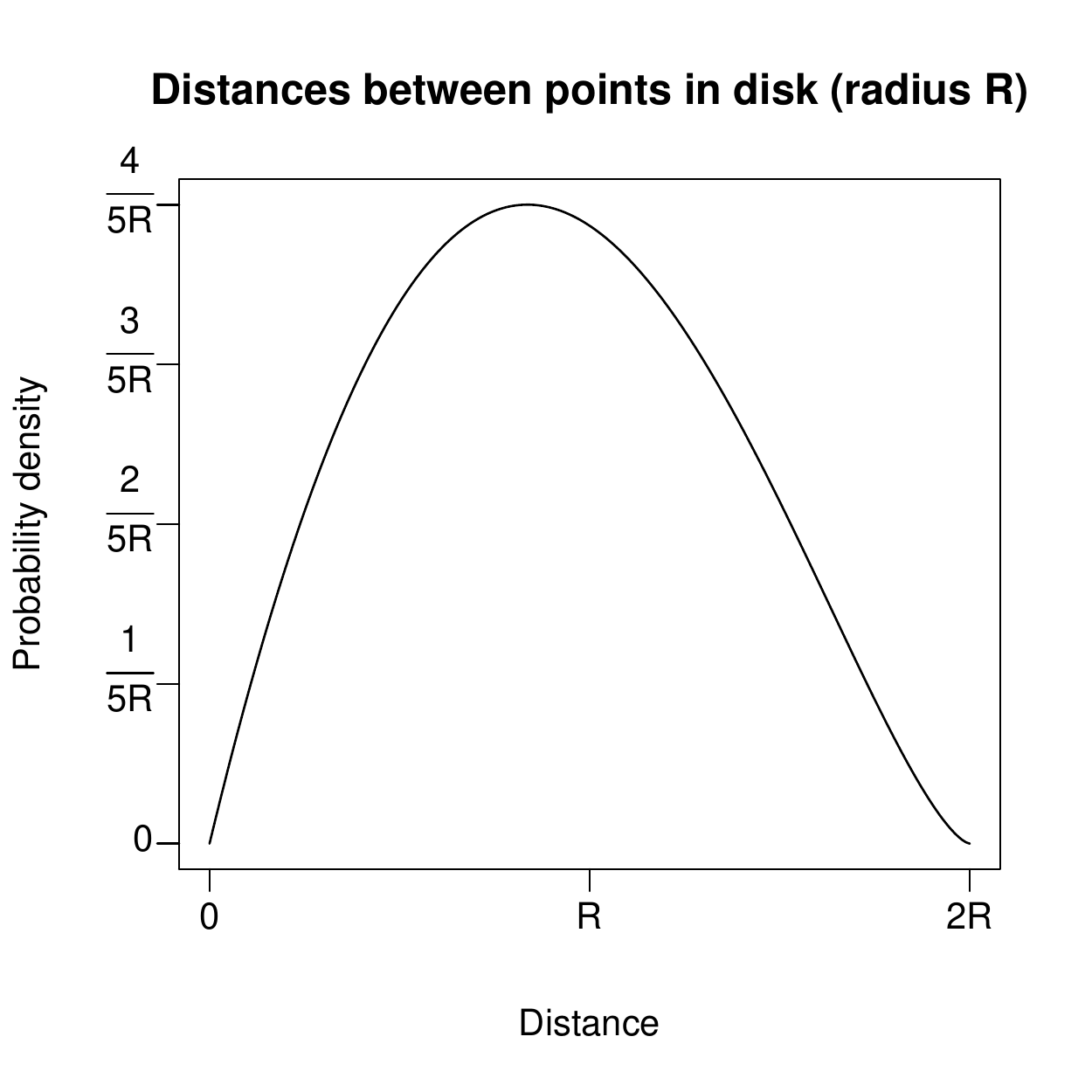}
\caption{The distribution of distances between points uniformly distributed on a disk of radius $R$.}
\label{fig:diskDistribution}
\end{figure}

Long-range correlations are especially important when calculating predictions of certain areal averages. With a `back of the envelope' calculation one can calculate the variance of a predicted spatial integral over a disk with radius $R$. Let $\hat{ r }(d)$ be the estimated covariance, and let $ r (d)$ be the true covariance such that, 
$$ \hat{ r }(d) = r (d) + e(d), $$
so $e$ is the error in the covariance estimate a a function of distance.  Then if we denote the disk by $A$, and the true spatial field with $g(\mathbf{x})$, the variance of our spatial integral under the predictive distribution is: 
\begin{align*}
\widehat{\mbox{Var}}(g(A)) &=  \int_A \int_A \widehat{\mbox{Cov}}(\mathbf{u}, \mathbf{v}) \ d\mathbf{u} \ d\mathbf{v} \\
&=  \int_A \int_A r (||\mathbf{u} - \mathbf{v} || ) + e(||\mathbf{u} - \mathbf{v} || ) \ d\mathbf{u} \ d\mathbf{v} \\
&= \mbox{Var}(g(A)) + \int_A \int_A e(||\mathbf{u} - \mathbf{v} || ) \ d\mathbf{u} \ d\mathbf{v}.
\end{align*}
Let $D$ be the random distance between any two points chosen in the disk with independent uniform distributions. Then \citet{tuckwell2018elementary} shows the density of $D$ is: 
$$ p_D(d) = \begin{cases}
\frac{4d}{ \pi R^2}  \left ( \arccos \left ( \frac{d}{2R}\right ) - \frac{d}{2R}\sqrt{1 -  \left ( \frac{d}{2R}  \right )^2} \right ), & 0 \leq d \leq 2R \\
0, & \text{otherwise.}
\end{cases}$$
Hence, 
\begin{align*}
\widehat{\mbox{Var}}(g(A)) &=  \mbox{Var}(g(A)) + \int_0^{2R} e( D ) \cdot \frac{4d}{ \pi R^2} \left ( \arccos \left ( \frac{d}{2R}\right ) - \frac{d}{2R}\sqrt{1 -  \left ( \frac{d}{2R}  \right )^2} \right ) \ d D.
\end{align*}
Fig. \ref{fig:diskDistribution} shows that the density $p_D(d)$ roughly parabolic with peak just under $R$ (approximately $0.834 R$), and has zeros at 0 and $2R$. Because of this, errors in very short and very long-range correlations are less relevant than errors in the assumed correlation function at the spatial scale near the radius of the area over which we integrate, $R$, when calculating predictive uncertainties. This is of course not the full story, since the covariance structure conditional on the data will not be so neatly stationary and isotropic, and will likely have shorter spatial range. At the same time, we believe this shows greater emphasis must be placed on long range spatial correlations when producing area level predictions, especially in large areas.

 \FloatBarrier
 
\section*{Appendix B: ELK Sparse Matrix Computations}
\label{sec:computation}

The computational performance of our implementation of ELK within \textit{inla} is almost entirely determined by how quickly the sparse precision matrix of the basic coefficients $\mathbf{c}$ can be generated. As such, we precompute any information for this task that will improve the performance. Recall that thus basis coefficients for each layer follow independent SAR models with mean zero Gaussian distribution, $\mathbf{c}_l \sim \mbox{MVN}(\mathbf{0},  \alpha_l \sigma_{\mathrm{S}}^2 \mathbf{B}^{-1}_l \mathbf{B}^{-T}_l)$, with, 
$$ \mathbf{B}_{l,i,j} = \begin{cases}
4 + \kappa_l^2, & i=j \\
-1, & i \in N_l(j) \\
0, & \text{otherwise,}
\end{cases}$$
where $N_l(j)$ is this set of indices of lattice knots in layer $l$ neighboring lattice knot $i$. The precision matrix for layer $l$, $\mathbf{Q}_l$, can therefore be represented as, 

$$\mathbf{Q}_l =  \frac{ \omega_l}{ \alpha_l \sigma_{\mathrm{S}}^2 }  \left  (\kappa_l^4 \mathbf{I}_{m(l)} - \kappa^2_l (\mathbf{D}^l + (\mathbf{D}^l)^T) + (\mathbf{D}^l)^T \mathbf{D}^l \right ), $$
\noindent
for matrices, 
\begin{align*}
\mathbf{D}^l &= \mathbf{D}^l_x + \mathbf{D}^l_y\\
\mathbf{D}^l_x &= \mathbf{I}_{m_y(l)} \otimes \boldsymbol{\nabla}_{m_x(l)}^2 \\
\mathbf{D}^l_y &= \mathbf{I}_{m_x(l)} \otimes \boldsymbol{\nabla}_{m_y(l)}^2,
\end{align*}
where $m_x(l)$ and $m_y(l)$ are the number of basis functions in the horizontal and vertical directions of layer $l$, $\mathbf{I}_{m_x(l)}$ and $\mathbf{I}_{m_y(l)}$ are $m_x(l) \times m_x(l)$ and $m_y(l) \times m_y(l)$ identity matrices respectively, and `$\otimes$' denotes the Kronecker product. Note that the variance normalization factor $ \omega_l$ is a function of $ \kappa_l$, although we leave out this dependence in the notation for simplicity. We can therefore precompute $\mathbf{D}^l + (\mathbf{D}^l)^T$ and $\mathbf{D}^l)^T \mathbf{D}^l $ in order to calculate $\mathbf{Q}_l $ as quickly as possible for each chosen value of $ \kappa_l$.

Since there is no exact closed form solution for the functions $f_l:  \kappa_l \mapsto  \omega_l, \ l=1,\ldots,L $, they are approximated using monotonic smoothing splines \citep{hyman1983accurate} fit on a log-log scale over a set of reasonable effective ranges for each layer.  Throughout this paper, the effective ranges used for fitting $f_1$ vary from a fifth of the first layer lattice width to the diameter of the spatial domain, and the effective ranges used when fitting subsequent $f_l$ shrink proportionally with the corresponding lattice widths. Hence, if $w$ is the diameter of the spatial domain, then each $f_l$ is fit with effective ranges varying in the interval \(\left (\frac{\delta_l}{5}, \ \frac{\delta_1}{ \delta_l} \cdot \frac{w}{5} \right )\). We find the splines are nearly linear, so estimates of $f_l$ are very accurate even somewhat outside of the interval used for fitting.

\end{document}


\def\spacingset#1{\renewcommand{\baselinestretch}%
{#1}\small\normalsize} \spacingset{1}


\if0\blind
{
  \title{Supplementary Materials for Bayesian Multiresolution Modeling of Georeferenced Data}
  \author{John Paige \thanks{
    John Paige was supported by The National Science Foundation Graduate Research Fellowship Program under award DGE-1256082, and Jon Wakefield was supported by the National Institutes of Health under award R01CAO95994.}\hspace{.2cm}\\
    Department of Statistics, University of Washington,\\
    Geir-Arne Fuglstad \\
    Department of Mathematical Sciences, NTNU, \\
    Andrea Riebler \\
    Department of Mathematical Sciences, NTNU, \\
    and Jon Wakefield \\
    Departments of Statistics and Biostatistics, University of Washington}
  \maketitle
} \fi

\if1\blind
{
  \bigskip
  \bigskip
  \bigskip
  \begin{center}
    {\LARGE\bf Supplementary Materials for Bayesian Multiresolution Modeling of Georeferenced Data}
\end{center}
  \medskip
} \fi

\bigskip

\appendix
\renewcommand{\thesection}{S.\arabic{section}}

\section{Fuzzy Coverage and Interval Width for Count Data}
\label{sec:fuzzyCoverage}

For observation $i$, let $Q_{\alpha/2}^i$ and $Q_{1-\alpha/2}^i$ be the discrete $\alpha/1$ and $1-\alpha$ quantiles of a predictive distribution for empirical proportion $y_i$ so that $p_l \equiv P(y_i< Q_{\alpha/2}^i)  \leq  \alpha  / 2$ and $p_u \equiv P(y_i> Q_{1-\alpha/2}^i)  \leq  \alpha  / 2$. Then $P(Q_{\alpha/2}^i \leq y_i  \leq Q_{1-\alpha/2}^i)   \geq 1 -  \alpha $. We will show how to calculate coverage using fuzzy coverage intervals in order to achieve coverage closer to the nominal rate.

Rather than using a fixed uncertainty interval when performing hypothesis tests for discrete data, randomized tests involving randomized uncertainty intervals are the uniformly most powerful (UMP) one tailed and UMP unbiased (UMPU) two-tailed tests \citep[Chapters 3 and 4]{lehmann2005testing}. In a randomized test, we could randomly reject that $y_i$ is in our interval if $y_i$ is equal to $Q_{\alpha/2}^i$ or $Q_{1-\alpha/2}^i$ in such a way as to obtain equal tail rejection probabilities and achieve $1-\alpha$ coverage. In the lower tail case, we have:
\begin{align*}
 \alpha  / 2 &= P(\text{reject $y_i$ at lower tail}) \\
 &= P(y_i < Q_{\alpha/2}^i) + P(\text{reject $y_i$ at lower tail}, \ y_i = Q_{\alpha/2}^i).
\end{align*}
This implies we can choose $P(\text{reject $y_i$} \ \vert \ y_i = Q_{\alpha/2}^i)$ and $P(\text{reject $y_i$} \ \vert \ y_i = Q_{1-\alpha/2}^i)$ in the following way in order to achieve the correct coverage, assuming the predictive distribution is correct: 
\begin{align}
 \alpha  / 2  &= p_l +  P(\text{reject $y_i$} \ \vert \ y_i = Q_{\alpha/2}^i) \cdot P(y_i = Q_{\alpha/2}^i) \label{eq:lowerTail} \\
  \alpha  / 2  &= p_u +  P(\text{reject $y_i$} \ \vert \ y_i = Q_{1-\alpha/2}^i) \cdot P(y_i = Q_{1-\alpha/2}^i). \label{eq:upperTail}
\end{align}

Since the resulting coverage intervals are random, different statisticians may randomly report different results. To eliminate this possibility, we will follow the proposal of \citet{geyer2005fuzzy} to use fuzzy set theory to compute fuzzy intervals. To do this, we calculate the membership function for $U^i$, the \textit{fuzzy} uncertainty interval for the $i$th observation, as: 
\begin{equation}
I_{U^i}(y_i) = \begin{cases}
1, & Q_{\alpha/2}^i< y_i < Q_{1-\alpha/2}^i \\
P(\text{reject $y_i$} \ \vert \ y_i = Q_{\alpha/2}^i), & y_i = Q_{\alpha/2}^i \\
P(\text{reject $y_i$} \ \vert \ y_i = Q_{1-\alpha/2}^i), & y_i = Q_{1-\alpha/2}^i \\
0, & \text{otherwise,}
\end{cases}
\label{eq:fuzzy}
\end{equation}
where $P(\text{reject $y_i$} \ \vert \ y_i = Q_{\alpha/2}^i)$ and $P(\text{reject $y_i$} \ \vert \ y_i = Q_{\alpha/2}^i)$ are calculated from Eqs. (\ref{eq:lowerTail}-\ref{eq:upperTail}). We then calculate coverage for a single observation as the membership function of the fuzzy interval, 
$$ \mbox{Cvg}(y_i) = I_{U^i}(y_i), $$
in order to achieve the nominal coverage deterministically.

To calculate fuzzy credible interval width, we modify the standard width calculations by accounting for the rejection probabilities in Eqs. (\ref{eq:lowerTail}-\ref{eq:upperTail}): 
$$ \mbox{Width}(y_i) = Q_{1-\alpha/2}^i - Q_{\alpha/2}^i  - \frac{1}{N_i} \left [ P(\text{reject $y_i$} \ \vert \ y_i = Q_{\alpha/2}^i) + P(\text{reject $y_i$} \ \vert \ y_i = Q_{1-\alpha/2}^i) \right ], $$
where $\frac{1}{N_i}$ is the width of the discrete steps the interval width could increase by if fuzzy intervals were not used.

\section{Assessing Performance Under Multiscale Dependence: Additional Results}
\label{sec:simulationStudyAppendix}

In addition to the results for the simulation study shown in the main text, we also calculated scores for the predictions integrated over the nine grid cells throughout the domain, averaged over the 100 realizations for the single central grid cell without observations, and for the eight other grid cells containing observations. These average scores are respectively given in Tables \ref{tab:mixtureScoresAggregatedInner} and \ref{tab:mixtureScoresAggregatedOuter}. Figure \label{fig:exampleKnots} depicts the data domain and knots for the three lattice layer ELK model.

\begin{figure}
\centering
\image{width=4in}{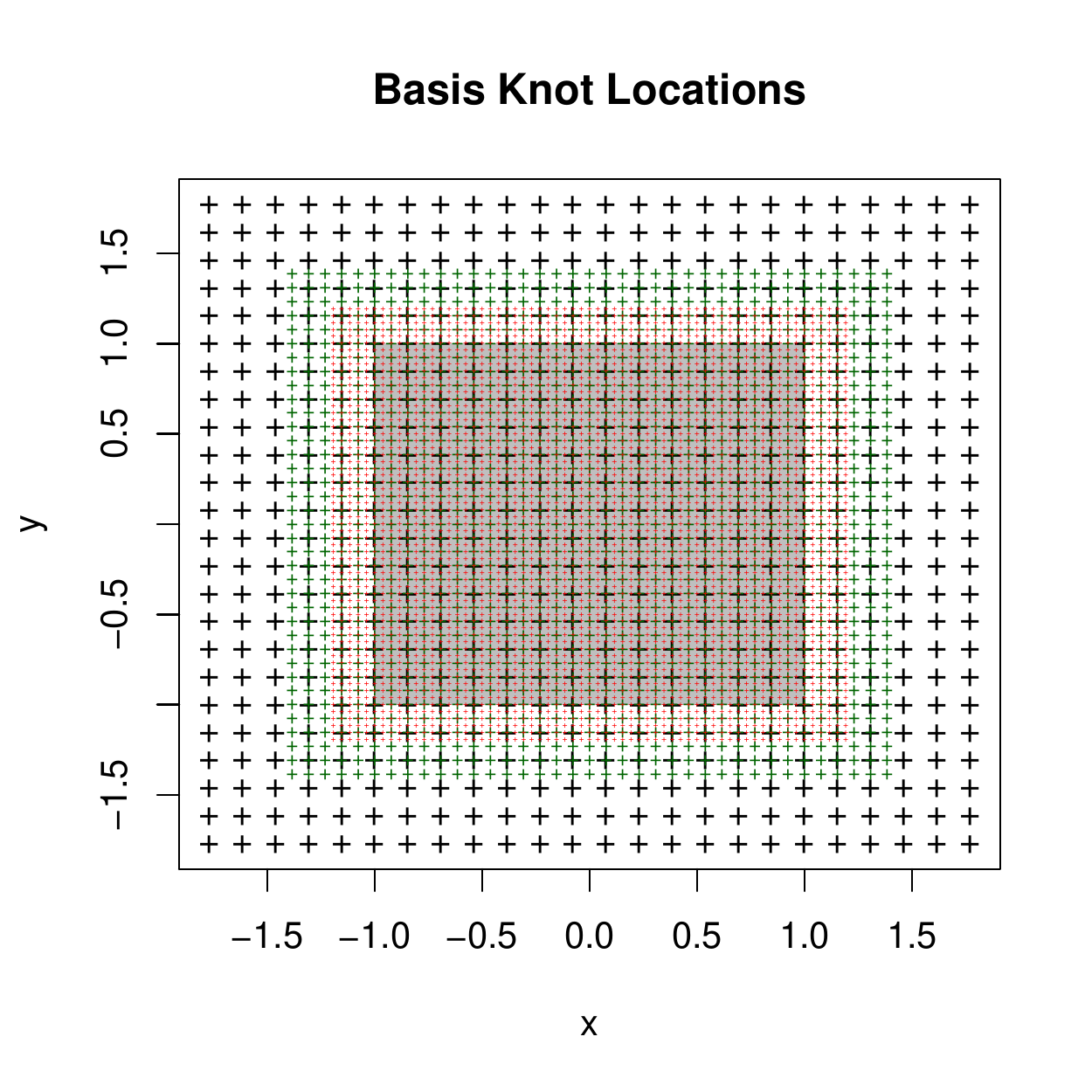}
\caption{An example of a set of three lattice layers for the given $[-1, 1] \times [-1, 1]$ data domain represented by the shaded region. The `+' signs represent the knot points where the basis functions are centered: large black symbols representing the coarsest, first layer, medium green symbols representing the second layer, and small red symbols representing the last, finest layer.}
\label{fig:exampleKnots}
\end{figure}

\begin{table}
\centering
\begin{tabular}{rllll}
  \hline  
 & RMSE & CRPS & 80\% Cvg & CI Width \\ 
  \hline    \hline
  SPDE & \textit{0.37} & \textit{0.21} & \textit{60} & \textbf{0.61} \\ 
  LK & 0.31 & 0.18 & 75 & 0.68 \\ 
  ELK-F & 0.33 & 0.19 & 67 & 0.64 \\ 
  ELK-T & \textbf{0.27} & \textbf{0.16} & \textbf{76} & 0.66 \\ 
    \hline
\end{tabular}
\caption{Scoring rules for predictions of integrals of the latent field over the center most of the nine cells in the $3 \times 3$ regular grid. The scoring rules are averaged over 100 simulated realizations and are calculated for each of the considered models. \textit{Italics} indicate worse performance, \textbf{boldface} indicates better performance.}
\label{tab:mixtureScoresAggregatedInner}
\end{table}

\begin{table}
\centering
\begin{tabular}{rllll}
  \hline  
 & RMSE & CRPS & 80\% Cvg & CI Width \\ 
  \hline    \hline
  SPDE & \textit{0.066} & \textit{0.037} & \textit{77} & \textit{0.16} \\ 
  LK & 0.062 & 0.035 & \textit{77} & \textbf{0.15} \\ 
  ELK-F & 0.063 & 0.035 & 78 & \textbf{0.15} \\ 
  ELK-T & \textbf{0.061} & \textbf{0.034} & \textbf{79} & \textbf{0.15} \\ 
    \hline
\end{tabular}
\caption{Scoring rules for predictions of integrals of the latent field over the outer most eight of the nine cells in the $3 \times 3$ regular grid. The scoring rules are averaged over 100 simulated realizations and are calculated for each of the considered models. \textit{Italics} indicate worse performance, \textbf{boldface} indicates better performance.}
\label{tab:mixtureScoresAggregatedOuter}
\end{table}

\section{Prevalence of Secondary Education in Kenya: \\ Survey Design and Additional Results}
\label{sec:applicationAppendix}

The KHDS follows a typical DHS design: it is a stratified, two-stage design, where the first stage consists of selecting enumeration areas (EAs) from each stratum with probability proportional to size (PPS) sampling, where the `size' used to calculate sampling probabilities is based on the number of households in each EA. The second stage consists of selecting 25 households randomly within each EA, \citep{KDHS2014,samplingManualDHS}. Strata are based on the 47 counties crossed with official urban/rural designations, where two counties, Nairobi and Mombasa, are entirely urban, making 92 strata in total. 1,612 clusters are sampled from the 96,251 EAs in Kenya that are based on the 2009 Kenya Population and Housing Census \citepalias{KenyaCensus:2009}.

\begin{sidewaysfigure}
\centering
\includegraphics[width=\textwidth]{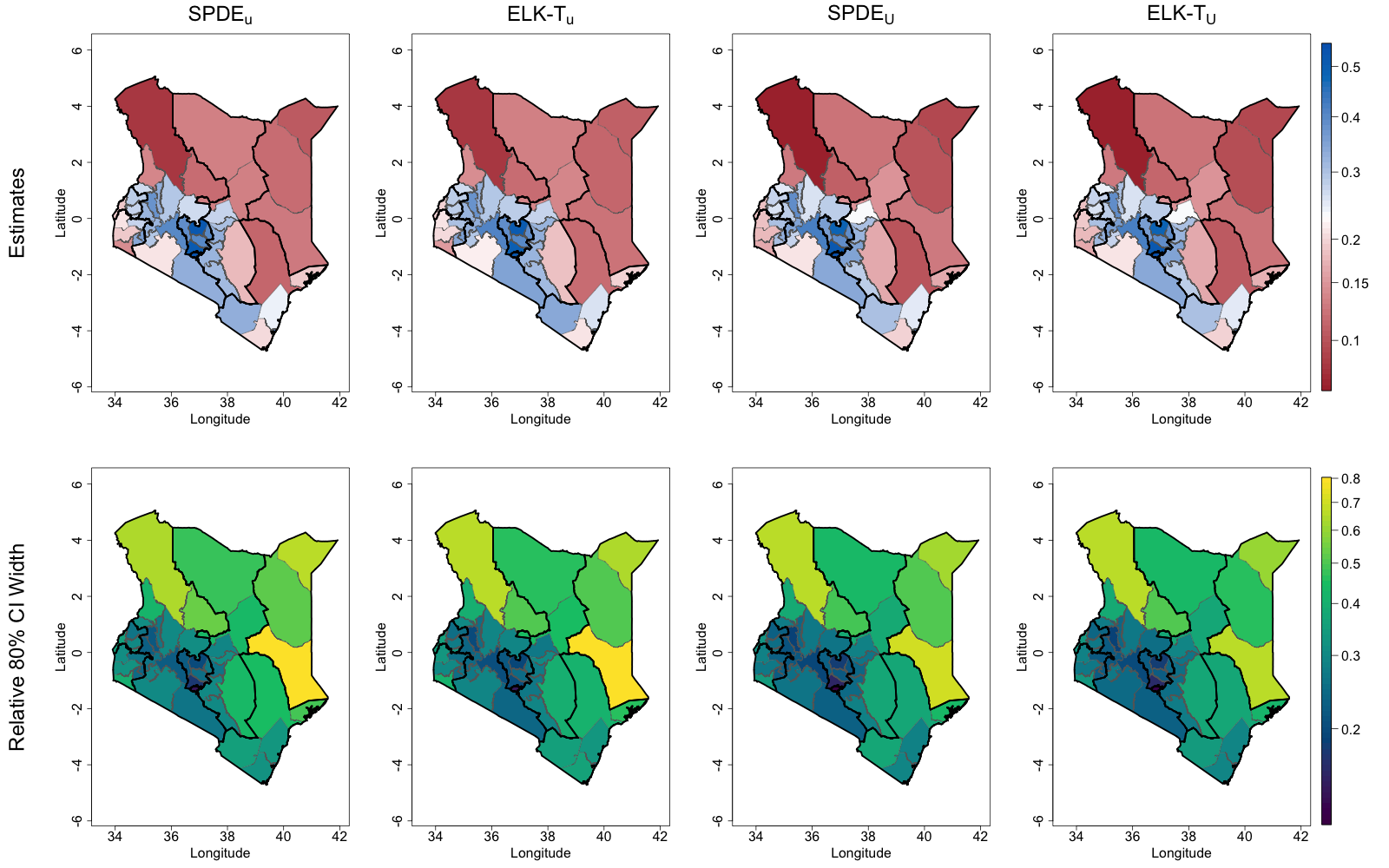}
\caption{Central county level predictions (top row) and relative 80\% credible interval widths (bottom row) of secondary education prevalence for young women in Kenya in 2014. Models with subscript `u' and `U' respectively do and do not include urban effects. Province and county borders are shown as black and grey lines respectively.}
\label{fig:countyPredictions}
\end{sidewaysfigure}

\begin{sidewaysfigure}
\centering
\includegraphics[width=\textwidth]{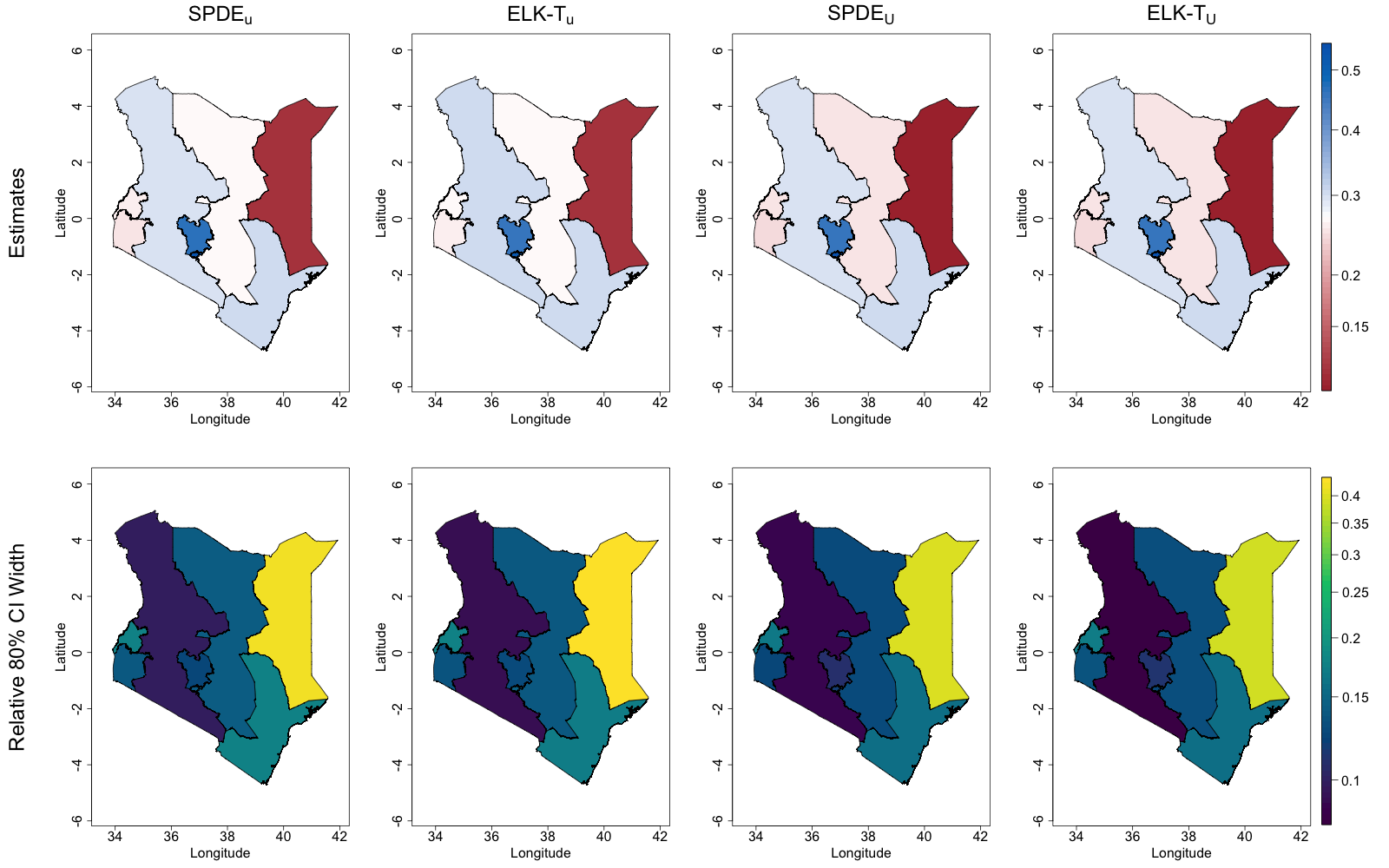}
\caption{Central province level predictions (top row) and relative 80\% credible interval widths (bottom row) of secondary education prevalence for young women in Kenya in 2014. Models with subscript `u' and `U' respectively do and do not include urban effects.}
\label{fig:regionPredictions}
\end{sidewaysfigure}

\begin{table}
\centering
\begin{tabular}{lrrrrr}
\toprule
Parameter & Est & SD & Q10 & Q50 & Q90\\
\midrule
\addlinespace[0.3em]
\multicolumn{6}{l}{\textit{\textbf{SPDE\textsubscript{U}}}}\\
\hspace{1em}Intercept & -2.392 & 0.262 & -2.381 & -2.724 & -2.073\\
\hspace{1em}Urban & 0.927 & 0.067 & 0.927 & 0.841 & 1.012\\
\hspace{1em}Total Var & 1.244 & 0.233 & 0.941 & 1.194 & 1.609\\
\hspace{1em}Spatial Var & 0.824 & 0.227 & 0.555 & 0.775 & 1.167\\
\hspace{1em}Cluster Var & 0.421 & 0.052 & 0.342 & 0.419 & 0.510\\
\hspace{1em}Total SD & 1.111 & 0.103 & 0.970 & 1.092 & 1.268\\
\hspace{1em}Spatial SD & 0.899 & 0.124 & 0.745 & 0.880 & 1.080\\
\hspace{1em}Cluster SD & 0.647 & 0.040 & 0.585 & 0.647 & 0.714\\
\hspace{1em}Range (km) & 203 & 43 & 154 & 196 & 267\\
\addlinespace[0.3em]
\multicolumn{6}{l}{\textit{\textbf{ELK-T\textsubscript{U}}}}\\
\hspace{1em}Intercept & -2.433 & 0.316 & -2.836 & -2.416 & -2.055\\
\hspace{1em}Urban & 0.930 & 0.067 & 0.844 & 0.930 & 1.016\\
\hspace{1em}Total Var & 1.298 & 0.290 & 0.977 & 1.206 & 1.705\\
\hspace{1em}Spatial Var & 0.870 & 0.277 & 0.556 & 0.780 & 1.283\\
\hspace{1em}Cluster Var & 0.428 & 0.053 & 0.354 & 0.426 & 0.487\\
\hspace{1em}Total SD & 1.132 & 0.126 & 0.988 & 1.098 & 1.306\\
\hspace{1em}Spatial SD & 0.921 & 0.146 & 0.746 & 0.883 & 1.133\\
\hspace{1em}Cluster SD & 0.653 & 0.041 & 0.595 & 0.653 & 0.698\\
\hspace{1em}Range\textsubscript{1} (km) & 571 & 451 & 315 & 375 & 1365\\
\hspace{1em}Range\textsubscript{2} (km) & 135 & 51 & 78 & 118 & 199\\
\hspace{1em}$ \alpha_1$ & 0.480 & 0.209 & 0.241 & 0.538 & 0.786\\
\hspace{1em}$ \alpha_2$ & 0.520 & 0.209 & 0.214 & 0.462 & 0.759\\
\bottomrule
\end{tabular}
\caption{Parameter posterior estimates, standard deviations, and 80\% CIs for the given models fit to secondary education completion KDHS data for women aged 20-29 in Kenya in 2014.}
\label{tab:parameters}
\end{table}

\begin{table}
\centering
\begin{tabular}{lrrrrrr}
\toprule
\multicolumn{1}{c}{\em{\textbf{ }}} & \multicolumn{3}{c}{\em{\textbf{SPDE\textsubscript{U}}}} & \multicolumn{3}{c}{\em{\textbf{ELK-T\textsubscript{U}}}} \\
\cmidrule(l{3pt}r{3pt}){2-4} \cmidrule(l{3pt}r{3pt}){5-7}
Province & Est & Q10 & Q90 & Est & Q10 & Q90\\
\midrule
Central & 0.4578 & 0.4331 & 0.4826 & 0.4566 & 0.4326 & 0.4813\\
Coast & 0.3030 & 0.2791 & 0.3276 & 0.3047 & 0.2814 & 0.3287\\
Eastern & 0.2542 & 0.2377 & 0.2706 & 0.2556 & 0.2381 & 0.2734\\
Nairobi & 0.5403 & 0.5082 & 0.5727 & 0.5391 & 0.5066 & 0.5706\\
North Eastern & 0.1059 & 0.0856 & 0.1281 & 0.1046 & 0.0845 & 0.1261\\
Nyanza & 0.2429 & 0.2277 & 0.2578 & 0.2432 & 0.2278 & 0.2598\\
Rift Valley & 0.2935 & 0.2808 & 0.3066 & 0.2940 & 0.2809 & 0.3065\\
Western & 0.2492 & 0.2283 & 0.2696 & 0.2489 & 0.2291 & 0.2684\\
\bottomrule
\end{tabular}
\caption{\label{tab:predictionsRegion} Province predictions and 80\% CIs for prevalence of secondary education completion for women aged 20-29 in Kenya in 2014.}
\end{table}

\FloatBarrier

\begin{longtable}{lrrrrrr}
\caption{\label{tab:predictionsCounty} County level predictions and 80\% CIs for prevalence of secondary education completion for women aged 20-29 in Kenya in 2014.}\\
\toprule
\multicolumn{1}{c}{\em{\textbf{ }}} & \multicolumn{3}{c}{\em{\textbf{SPDE\textsubscript{U}}}} & \multicolumn{3}{c}{\em{\textbf{ELK-T\textsubscript{U}}}} \\
\cmidrule(l{3pt}r{3pt}){2-4} \cmidrule(l{3pt}r{3pt}){5-7}
County & Est & Q10 & Q90 & Est & Q10 & Q90\\
\midrule
\endfirsthead
\caption[]{County level predictions and 80\% CIs for prevalence of secondary education completion for women aged 20-29 in Kenya in 2014. \textit{(continued)}}\\
\toprule
\multicolumn{1}{c}{\em{\textbf{ }}} & \multicolumn{3}{c}{\em{\textbf{SPDE\textsubscript{U}}}} & \multicolumn{3}{c}{\em{\textbf{ELK-T\textsubscript{U}}}} \\
\cmidrule(l{3pt}r{3pt}){2-4} \cmidrule(l{3pt}r{3pt}){5-7}
County & Est & Q10 & Q90 & Est & Q10 & Q90\\
\midrule
\endhead
\
\endfoot
\bottomrule
\endlastfoot
Baringo & 0.2615 & 0.2267 & 0.2977 & 0.2607 & 0.2256 & 0.2983\\
Bomet & 0.2768 & 0.2418 & 0.3128 & 0.2765 & 0.2408 & 0.3131\\
Bungoma & 0.2734 & 0.2422 & 0.3063 & 0.2748 & 0.2440 & 0.3082\\
Busia & 0.1810 & 0.1513 & 0.2103 & 0.1806 & 0.1524 & 0.2122\\
Elgeyo Marakwet & 0.2935 & 0.2591 & 0.3292 & 0.2953 & 0.2609 & 0.3292\\
\addlinespace
Embu & 0.3183 & 0.2800 & 0.3568 & 0.3205 & 0.2829 & 0.3590\\
Garissa & 0.1250 & 0.0829 & 0.1719 & 0.1225 & 0.0824 & 0.1692\\
Homa Bay & 0.1795 & 0.1552 & 0.2032 & 0.1803 & 0.1558 & 0.2067\\
Isiolo & 0.1467 & 0.1195 & 0.1757 & 0.1483 & 0.1216 & 0.1765\\
Kajiado & 0.3372 & 0.2972 & 0.3755 & 0.3394 & 0.3012 & 0.3770\\
\addlinespace
Kakamega & 0.2573 & 0.2263 & 0.2916 & 0.2560 & 0.2248 & 0.2882\\
Kericho & 0.3510 & 0.3127 & 0.3891 & 0.3513 & 0.3142 & 0.3889\\
Kiambu & 0.5182 & 0.4776 & 0.5577 & 0.5164 & 0.4742 & 0.5565\\
Kilifi & 0.2461 & 0.2113 & 0.2816 & 0.2471 & 0.2119 & 0.2841\\
Kirinyaga & 0.3883 & 0.3461 & 0.4285 & 0.3879 & 0.3458 & 0.4331\\
\addlinespace
Kisii & 0.3039 & 0.2715 & 0.3396 & 0.3055 & 0.2709 & 0.3417\\
Kisumu & 0.3057 & 0.2714 & 0.3416 & 0.3041 & 0.2705 & 0.3384\\
Kitui & 0.1631 & 0.1335 & 0.1942 & 0.1652 & 0.1352 & 0.1958\\
Kwale & 0.1978 & 0.1696 & 0.2278 & 0.1999 & 0.1726 & 0.2298\\
Laikipia & 0.2856 & 0.2480 & 0.3227 & 0.2862 & 0.2519 & 0.3237\\
\addlinespace
Lamu & 0.1647 & 0.1297 & 0.2011 & 0.1684 & 0.1317 & 0.2102\\
Machakos & 0.3514 & 0.3086 & 0.3951 & 0.3515 & 0.3073 & 0.3977\\
Makueni & 0.2821 & 0.2421 & 0.3226 & 0.2845 & 0.2448 & 0.3267\\
Mandera & 0.0945 & 0.0659 & 0.1242 & 0.0943 & 0.0655 & 0.1256\\
Marsabit & 0.1238 & 0.0981 & 0.1520 & 0.1223 & 0.0955 & 0.1514\\
\addlinespace
Meru & 0.2330 & 0.2001 & 0.2661 & 0.2345 & 0.2016 & 0.2682\\
Migori & 0.1676 & 0.1398 & 0.1970 & 0.1689 & 0.1399 & 0.1973\\
Mombasa & 0.4215 & 0.3761 & 0.4680 & 0.4241 & 0.3809 & 0.4665\\
Murang'a & 0.3897 & 0.3465 & 0.4326 & 0.3859 & 0.3419 & 0.4289\\
Nairobi & 0.5403 & 0.5082 & 0.5727 & 0.5391 & 0.5066 & 0.5706\\
\addlinespace
Nakuru & 0.4185 & 0.3780 & 0.4587 & 0.4196 & 0.3797 & 0.4607\\
Nandi & 0.2799 & 0.2504 & 0.3097 & 0.2777 & 0.2487 & 0.3093\\
Narok & 0.2091 & 0.1817 & 0.2364 & 0.2094 & 0.1837 & 0.2370\\
Nyamira & 0.3278 & 0.2894 & 0.3658 & 0.3267 & 0.2855 & 0.3659\\
Nyandarua & 0.3586 & 0.3187 & 0.4004 & 0.3599 & 0.3208 & 0.3986\\
\addlinespace
Nyeri & 0.4863 & 0.4413 & 0.5294 & 0.4878 & 0.4427 & 0.5330\\
Samburu & 0.1101 & 0.0847 & 0.1383 & 0.1130 & 0.0831 & 0.1438\\
Siaya & 0.1871 & 0.1609 & 0.2149 & 0.1876 & 0.1618 & 0.2150\\
Taita Taveta & 0.3038 & 0.2509 & 0.3631 & 0.3040 & 0.2499 & 0.3599\\
Tana River & 0.1026 & 0.0845 & 0.1233 & 0.1030 & 0.0842 & 0.1230\\
\addlinespace
Tharaka-Nithi & 0.3024 & 0.2638 & 0.3435 & 0.3037 & 0.2643 & 0.3428\\
Trans-Nzoia & 0.2465 & 0.2136 & 0.2826 & 0.2473 & 0.2137 & 0.2810\\
Turkana & 0.0707 & 0.0495 & 0.0965 & 0.0714 & 0.0504 & 0.0963\\
Uasin Gishu & 0.3883 & 0.3516 & 0.4239 & 0.3885 & 0.3506 & 0.4248\\
Vihiga & 0.2699 & 0.2341 & 0.3053 & 0.2694 & 0.2347 & 0.3060\\
\addlinespace
Wajir & 0.0996 & 0.0755 & 0.1254 & 0.0982 & 0.0745 & 0.1241\\
West Pokot & 0.1278 & 0.1032 & 0.1525 & 0.1291 & 0.1052 & 0.1551\\*
\end{longtable}


 \FloatBarrier

\bibliographystyle{apalike}

\bibliography{myBib}